\documentclass[aps,prx,twocolumn,amsmath,nofootinbib,amssymb,superscriptaddress,longbibliography,10pt]{revtex4}
\usepackage{graphicx}
\usepackage{ulem}
\usepackage{soul}
\usepackage{float} 
\usepackage[colorlinks, linkcolor=red , citecolor= blue]{hyperref}
\usepackage{breakurl}

\usepackage{verbatim}
\usepackage{esint}
\usepackage{marginnote}

\newcommand{\RNum}[1]{\uppercase\expandafter{\romannumeral #1\relax}}

\def \beq {\begin{eqnarray}}
\def \eeq {\end{eqnarray}}

\begin{document}

\title{Strongly coupled quantum phonon fluid in a solvable model}

\author{Evyatar Tulipman}
\affiliation{Department of Condensed Matter Physics, Weizmann Institute of Science, Rehovot, 76100, Israel}
\author{Erez Berg}
\affiliation{Department of Condensed Matter Physics, Weizmann Institute of Science, Rehovot, 76100, Israel}
\begin{abstract}
We study a model of a large number of strongly coupled phonons that can be viewed as a bosonic variant of the Sachdev-Ye-Kitaev model. We determine the phase diagram of the model which consists of a glass phase and a disordered phase, with a first-order phase transition separating them. We compute the specific heat of the disordered phase, with which we diagnose the high-temperature crossover to the classical limit. We further study the real-time dynamics of the disordered phase, where we identify three dynamical regimes as a function of temperature. Low temperatures are associated with a semiclassical regime, where the phonons can be described as long-lived normal modes. High temperatures are associated with the classical limit of the model. For a large region in parameter space, we identify an intermediate-temperatures regime, where the phonon lifetime is of the order of the Planckian time scale $\hbar/k_B T$.
\end{abstract}
\maketitle
\section{Introduction}
\label{intro}

Is there a fundamental limit to how fast can a quantum many-body system relax back to equilibrium? This basic question arises frequently in the interpretation of experiments in condensed matter systems \cite{zaanen_why_2004,sachdev_quantum_2011,bruin_similarity_2013,legros_universal_2019,cao_strange_2020,zhang_anomalous_2017}. In particular, transport coefficients can be related to relaxation times of electrical and thermal currents \cite{davison_holographic_2014,blake_quantum_2018,patel_quantum_2017,patel_magnetotransport_2018,davison_thermoelectric_2017,chowdhury_translationally_2018}. Often, the current relaxation times are tied to extrinsic mechanisms, such as the momentum loss rate due to impurity scattering. However, in setups where the bottleneck for current relaxation is the intrinsic thermalization time in the system, it has been proposed that the relaxation time has to obey a fundamental `Planckian' bound, $\tau_{th} \geq \alpha \frac{\hbar}{k_B T}$, where $\alpha$ is an unknown constant of order unity \cite{hartnoll_theory_2015,Nussinov_2020}. 

Evidence for this intriguing idea comes both from solvable models, such as holographic theories and systems near quantum critical points \cite{kovtun_viscosity_2005,shenker_black_2014,roberts_localized_2015,sachdev_quantum_2011,blake_universal_2016,hartnoll_holographic_2018,Gu17,zaanen_liu_sun_schalm_2015,geng_non-local_2020}, and from experiments \cite{bruin_similarity_2013,legros_universal_2019,cao_strange_2020}. Much attention has been devoted to electrical transport in `strange metals', where the Planckian bound is a natural way to explain the linear dependence of the resistivity on temperature. To test this hypothesis, the transport lifetime can be estimated using the Drude formula for the resistivity: $\rho = \frac{m^*}{n e^2 \tau}$, in materials where the electronic effective mass $m^*$ and density $n$ are well known. This procedure indeed yields $\tau \sim \frac{\hbar}{k_B T}$, with a coefficient of order one, in a host of different materials in regimes where $\rho \sim T$~\cite{bruin_similarity_2013,legros_universal_2019,cao_strange_2020}. 

Surprisingly, it was recently noted that in a wide class of insulating compounds at high temperature, similar physics may be at play \cite{behnia_lower_2019,martelli_thermal_2018,zhang_thermalization_2019}. In these systems (for example, complex oxides like SrTiO$_3$), the thermal current is carried by lattice vibrations. Around room temperature and above, the thermal conductivity $\kappa$ is approximately inversely proportional to temperature \cite{Ziman_2001}. Defining the thermal transport lifetime as $\tau = D_{th}/v^2_{ph}$, where $D_{th}$ is the thermal diffusivity (directly measured or obtained via the Einstein relation $\kappa = c D_{th}$, where $c$ is the specific heat) and $v_{ph}$ is a characteristic phonon velocity, operationally defined as the averaged speed of sound, gives again $\tau = \alpha\frac{\hbar}{ k_B T}$ where $\alpha$ is found to be in the range 1--3 in a host of different poorly thermally conducting materials. In contrast, in good thermal conductors $\alpha$ is much larger. For example, diamond (which exhibits $\kappa \sim 1/T$ over a range of temperatures explained as a result of phonon umklapp), $\alpha$ is found to be $\sim 50$ \cite{zhang_thermalization_2019}.  

This observation is particularly counter-intuitive since, at such elevated temperatures, one would naively expect the lattice dynamics to be essentially classical.
In a classical analysis of the lattice dynamics, any time scale must be proportional to $\sqrt{M}$, where $M$ is the ion mass. A simple dimensional analysis argument implies that $\tau_{\text{ph}} \sim \sqrt{M/m}$ where $m$ is the electron mass \cite{classical_scaling}. 
The observation of a short transport lifetime has the intriguing implication that in a broad temperature regime, the system should be thought of as a quantum-mechanical `fluid' of strongly coupled lattice vibrations, rather than in terms of individual long-lived phonon excitations. This view is supported by the fact that the estimated phonon mean free path is very short, of the order of the lattice spacing or less~\cite{zhang_thermalization_2019}. Furthermore, at least some of the complex oxide materials that show $\alpha\sim 1$ have very high frequency optical phonon branches, far exceeding room temperature, potentially explaining why the lattice dynamics is not fully classical even at room temperature and above. 

These intriguing observations call for a theoretical framework where the crossover from classical to quantum dynamics of lattice vibrations can be investigated. Considering a generic system of coupled non-linear oscillators, we expect that at sufficiently low temperatures, the system is always described in terms of low-energy, long-lived normal modes (or phonon quasi-particles). Conversely, at sufficiently high temperature, the dynamics is expected to become classical. The quantum relaxation time $\tau \sim \hbar/k_B T$ may thus appear only at intermediate temperature scales.\footnote{For an alternative interpretation of experiments in poor thermal conductors, where the apparent ``Planckian'' behavior originates from an interplay between the characteristic phonon speed and the relaxation time, see~\cite{mousatov_planckian_2019}.} This regime is the most difficult to analyze theoretically. 

\begin{figure}[t]
\centering
\includegraphics[width=\columnwidth]{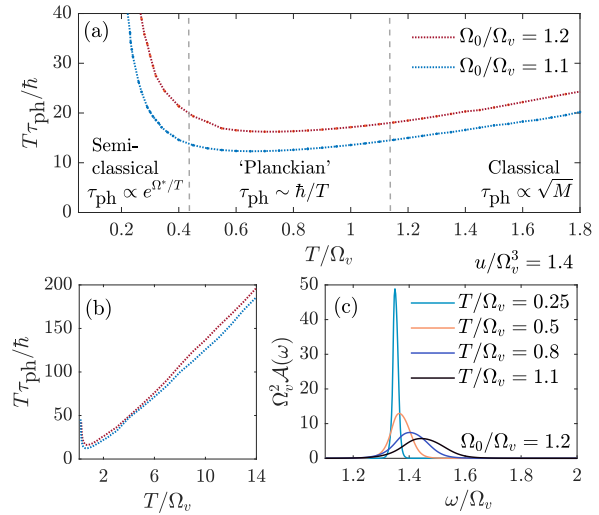}
\caption{ Dynamical properties of the disordered phase of the SB model. (a) shows the phonon lifetime in the units of the Planckian time scale $\hbar/T$ as a function of temperature for two values of $\Omega_0$. We identify three dynamical regimes of the model, a low temperature semi-classical regime, a high-temperature classical regime, and an intermediate `Planckian' regime (see Sec.~\ref{sec:DYNAMICS}), separated by the dashed gray lines in the figure. 
(b) presents the high-$T$ behavior of the phonon lifetime for the same set of parameters as in (a). This regime is associated with the approach to the classical limit, where the linearity of $T\tau_\text{ph} / \hbar$ suggests that the phonon lifetime becomes independent of $T$ (see Eqn.~\ref{eq:tau_high_T}). (c) shows the phonon spectral function $\mathcal{A}(\omega)$ for four increasing temperatures. The dynamical crossover from the semiclassical regime (with $T/\Omega_v = 0.25$) to the `Planckian' regime is demonstrated by the significant broadening of the spectral peak. Here, $u/\Omega_v^3 = 1.4$, and $\Omega^*$ denotes the $T\to0$ renormalized phonon frequency (see Eqn.~\ref{eq:tau_low_T}).
}
\label{fig:single_band_demo}
\end{figure}

In this work, we propose a simple model of strongly interacting phonons, that can be used to address the above questions. The model can be viewed as a bosonic variant of the Sachdev-Ye-Kitaev (SYK) model \cite{sachdev_gapless_1993,Kitaev_SYK_talk,maldacena_remarks_2016} of $N$ degrees of freedom coupled via an all-to-all, random interaction; similarly to SYK, the model is solvable in the large $N$ limit. The real-time dynamics of the model is found to follow the trends described above, with long relaxation times associated with long-lived phonon modes at low $T$, a crossover to classical nonlinear dynamics with $\tau \sim \sqrt{M}$ at high $T$, and a broad intermediate $T$ regime where the lifetime is of the order of the Planckian time scale $\hbar/k_B T$. Some representative results for the phonon lifetime $\tau_\text{ph}$ scaled by $\hbar/k_B T$ as a function of temperature are shown in Fig.~\hyperref[fig:single_band_demo]{\ref{fig:single_band_demo}a,b}.

This paper is organized as follows. In Section~\ref{sec:model} we introduce the model and identify the relevant energy scales. In Section~\ref{sec:Thermodynamics} we discuss some of its thermodynamic properties. We map out the phase diagram of the model, and in addition we discuss the specific heat in the disordered phase. The dynamics of the model, and in particular the phonon lifetime as a probe to identify different dynamical regimes, are discussed in Section~\ref{sec:DYNAMICS}. In Section~\ref{sec:MB_model} we discuss a generalized version of the model with multiple phonon branches.  In Section~\ref{relation_to_syk_main_text} we comment on several differences between the fermionic SYK model and its bosonic variant. Details on the imaginary- and real-time derivations are given respectively in Appendices \ref{sec:replica_formalism} and \ref{sec:keldysh_formalism}, and Appendix \ref{sec:numerical_methods} contains details on the numerical methods.

\section{Model}
\label{sec:model}

We consider a system of $N$ coupled nonlinear oscillators in $d=0+1$ space-time dimensions, described by the Hamiltonian
\begin{eqnarray}
H &=& \sum_{i=1}^N \frac{\pi_i^2}{2M} + \frac{M\Omega_i^2}{2}  \phi_i^2 + \frac{1}{N}\sum_{i,j,k} \tilde{v}_{ijk} \phi_i \phi_j \phi_k \nonumber \\ 
&+& \frac{\tilde{u}}{4N} \left( \sum_{i=1}^N \phi_i^2 \right)^2. 
\label{eq:H}
\end{eqnarray}
Here, $\phi_i$ is the displacement of the $i$th mode, $\pi_i$ is the conjugate momentum (such that $[\phi_i,\pi_j] = i\hbar \delta_{ij}$), $M$ is the mass, and $\Omega_i$ is the frequency in the absence of non-linearity. We will begin with the case where the frequencies are all the same, $\Omega_i = \Omega_0$, considering a more general situation later. The cubic couplings $\tilde{v}_{ijk}$ are chosen to be independent random Gaussian variables, each satisfying $\overline{\tilde{v}_{ijk}}=0$ and $\overline{\tilde{v}_{ijk}^2} = 2\tilde{v}^2$ (no sum), where $\overline{(\cdot)}$ denotes averaging over realizations of $\tilde{v}_{ijk}$. The quartic interaction $\tilde{u}>0$ stabilizes the system (when $\tilde{u}=0$, the energy is not bounded from below, due to the cubic term). The Hamiltonian in Eqn.~\ref{eq:H} is similar to the spherical p-spin model \cite{cugliandolo_quantum_2001}, studied in the context of quantum spin glasses; the difference is that in the p-spin model, $\frac{1}{N}\sum_i \phi_i^2$ is constrained to unity, whereas here this quantity is unconstrained (the $\tilde{u}$ term implements a `soft constraint' on the magnitude of  $\frac{1}{N}\sum_i \phi_i^{2}$). 

In addition to the energy scale $\hbar \Omega_0$, we can define two energy scales associated with the non-linear terms by dimensional analysis. The $\tilde{v}$ term defines an energy scale $\hbar \Omega_{v} = \left(\hbar^{2}/{M}\right)^{3/5}\tilde{v}^{2/5}$, while the $\tilde{u}$ term is associated with the scale $\hbar \Omega_{u} = \left(\hbar^{2}/M\right)^{2/3} \tilde{u}^{1/3}$. We set $\hbar=k_B = 1$ henceforth, unless stated otherwise. 
These energy scales become apparent in Lagrangian formulation under the rescaling $\phi_i\to \sqrt{M}\phi_i$,
\begin{eqnarray}
    \mathcal{L} &=& \frac{1}{2}\sum_{i=1}^N \left(\left(\partial_t\phi_i\right)^2 - \Omega_i^2  \phi_i^2\right) - \frac{1}{N}\sum_{i,j,k} v_{ijk} \phi_i \phi_j \phi_k \nonumber \\ 
    &-& \frac{u}{4N} \left( \sum_{i=1}^N \phi_i^2 \right)^2, 
\label{eq:L}
\end{eqnarray}
where the rescaled couplings are given by $u=\Omega_{u}^3$ and  $\overline{v_{ijk}^2}  \equiv 2v^2 = 2\Omega_{v}^5$, such that the energy scales of the system are given by $\Omega_0,v^{2/5}$ and $u^{1/3}$. In this work, we focus on the strong coupling regime of the model, where $\Omega_0\sim v^{2/5} \sim u^{1/3}$.  We use the rescaled Lagrangian formulation henceforth.


\section{Thermodynamics and Phase Diagram}
\label{sec:Thermodynamics}

In this section, we discuss some of the thermodynamic properties of the model. We study its phase diagram, and compute the specific heat of the disordered phase as a function of temperature. The main interest of this work is the existence of a `phonon fluid' regime - a dynamical crossover region in parameter space where phonons are not well-defined quasiparticles. However, in this region, it might be thermodynamically favorable for the system to realize an ordered (glassy) phase that masks the `phonon liquid' behavior. It is therefore crucial to map the phase diagram of the model and find the boundaries of the glass phase. In addition, we study the specific heat of the disordered phase that can serve as a simple thermodynamical diagnostic for the crossover from the quantum mechanical `phonon fluid' to the classical regime.

\subsection{Phase Diagram}

Let us consider the simple version of the model, where $\Omega_i=\Omega_0$ for all $i=1,...,N$, which we dub the `single-branch' (SB) model. We study its equilibrium phase diagram as a function of temperature $T$ and frequency $\Omega_0$, for fixed values of $v$, $u$ and $M$ that satisfy $\Omega_{v} \sim \Omega_{u}$.

We compute the disorder-averaged free-energy density within the framework of the replica method, where it is given by
\begin{equation}
    \beta \overline{f} = - \frac{1}{N}\overline{\ln{Z}} = - \frac{1}{N}\lim_{n\to0}\frac{\overline{Z^n}-1}{n}.
    \label{eq:replica_trick}
\end{equation}
Here, $f$ is the free-energy per mode, $\beta$ is the inverse temperature and $Z^n$ is the replicated partition function of $n\in\mathbb{N}$ replicas of the model, where $n$ is then analytically continued to zero.
We proceed by introducing bilocal fields $G_{\alpha\beta}\left(
\tau,\tau'\right) = \frac{1}{N}\sum_i \phi_i^{\alpha}\left(
\tau\right)\phi_i^{\beta}\left(
\tau'\right)$ and enforce this identity with the Lagrange multiplier fields $\Pi_{\alpha\beta}\left(
\tau,\tau'\right)$, where $\alpha,\beta = 1,...,n$ are the replica indices and imaginary-time arguments are denoted by $\tau,\tau'\in\left[0,\beta\right]$. This allows us to express (see Appendix \ref{sec:replica_formalism}) the disorder-averaged replicated partition function as a functional integral of an effective action, $\overline{Z^n} = \int \mathcal{D}\boldsymbol{G} \mathcal{D}\boldsymbol{\Pi} \exp\left(-nNS_{\text{eff}}\right)$, where
\begin{widetext}
\begin{eqnarray}
\label{eq:S_eff_replicas}
 S_\text{eff}  &=& \frac{1}{2n}\ln\text{det}\left(\delta_{\alpha\beta}\delta(\tau-\tau')\left(-\partial_{\tau}^{2}+\Omega_{0}^2\right)-\Pi_{\alpha\beta}(\tau,\tau')\right) \\
    &-& \frac{1}{2n}\sum_{\alpha,\beta=1}^n\int_{0}^{\beta}d\tau d\tau'\left(\frac{v^{2}}{3}G_{\alpha\beta}\left(\tau,\tau'\right)^{3}-\frac{u}{2}G_{\alpha\beta}\left(\tau,\tau'\right)^{2}\delta_{\alpha\beta}\delta\left(\tau-\tau'\right)-\Pi_{\alpha\beta}\left(\tau,\tau'\right)G_{\alpha\beta}\left(\tau,\tau'\right)\right). \nonumber
\end{eqnarray}
\end{widetext}
In the limit of $N \to \infty$, this functional integral is controlled by the saddle point of the effective action, $\delta S_{\text{eff}}/\delta A_{\alpha\beta}=0$, $A=G,\Pi$, leading to a closed set of self-consistent equations. Different phases of the model are characterized by the replica-space structure of $G_{\alpha\beta}$. Namely, a diagonal $G_{\alpha\beta}$ corresponds to the disordered phase of the model, while a $G_{\alpha\beta}$ with non-zero off-diagonal elements corresponds to an ordered (glassy) phase, where the off-diagonal elements of $G_{\alpha\beta}$ are the order parameters. The saddle-point equations for the off-diagonal components of $G_{\alpha\beta}$ are identical to those of \cite{cugliandolo_quantum_2001}, which implies that only two stable solutions exists in replica space, the diagonal solution and the one-step replica symmetry breaking (1SRSB) solution.

The diagonal solution is given by $G_{\alpha\beta}\left(\tau,\tau'\right)\equiv G\left(\tau-\tau'\right)\delta_{\alpha\beta}$, and its corresponding self-consistent equations read
\begin{eqnarray}
    \hat{G}\left(i\omega_{k}\right) &=& \frac{1}{ \omega_{k}^{2} + \Omega_{0}^2   - \hat{\Pi}\left(i\omega_{k}\right)},
\nonumber \\
\Pi\left(\tau\right) &=& v^{2}G\left(\tau\right)^2-uG\left(\tau\right)\delta\left(\tau\right). 
\label{eq:G_Pi_diagonal}
\end{eqnarray}
Here, $\omega_k=2\pi k/\beta$, $k\in\mathbb{Z}$ are the bosonic Matsubara frequencies, hats denote Matsubara-frequency domain functions, and we have used the fact that $G$ is imaginary-time translationally invariant in thermal equilibrium.

The one-step replica symmetry breaking (1SRSB) solution is defined as ${G}_{\alpha\beta}\left( \tau,\tau' \right) \equiv \left( {g}_{d}\left( \tau-\tau' \right)-{g}_{EA} \right)\delta_{\alpha\beta}+\left({g}_{EA} - {g}_{0}\right)\epsilon_{\alpha\beta} + {g}_{0}$,
where $\epsilon_{\alpha\beta}=1$ if $\alpha$ and $\beta$ are in a diagonal block of size $m$ and $\epsilon_{\alpha\beta}=0$ otherwise, and $g_{EA}$ is the so-called Edwards-Anderson order parameter \cite{edwards_theory_1975,mezard_spin_1986}. The self-consistent equations and more details on the 1SRSB solution can be found in Appendix~\ref{sec:replica_formalism}. 

\begin{figure}[t]
\centering
\includegraphics[width=\columnwidth]{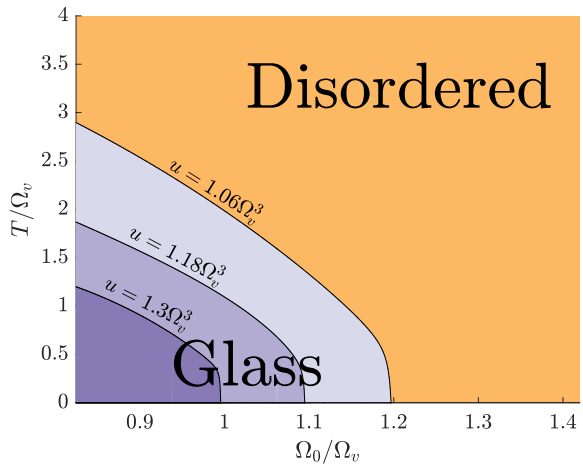}

\caption{Phase diagrams of the SB model for three values of $u$ as a function of $\Omega_0/\Omega_v$ and $T/\Omega_v$. The disordered phase corresponds the diagonal replica solution of the saddle-point equations and is denoted by the orange region. The glass phase corresponds to the one-step replica symmetry breaking solution of the saddle-point equations and is denoted by the purple regions, where different shades of purple corresponds to different values of $u$. Solid lines separating the glass and disordered phases correspond to a first order transition (for all values of $u$).} 
\label{fig:single_band_pd}
\end{figure}

One can solve these self-consistent equations numerically by an iterative procedure (see Appendix \ref{sec:numerical_methods}). Substituting the solutions back in $S_\text{eff}$ enables us to obtain the free-energy density of the two phases of the model and thereby to obtain its phase diagram. For a fixed $u$, we find that the model realizes two phases, a disordered (replica diagonal) phase, and a glass (1SRSB) phase. The transition between the two phases is first order. This transition is associated with a discontinuity in the order parameter $g_{EA}$ while the break-point parameter $m<1$.\footnote{This discontinuity might not be sufficient for a first-order transition if the break-point parameter $m\to1$ at the transition, because the effective number of degrees of freedom that are involved in the transition is $(1-m)g_{EA}$ \cite{cugliandolo_quantum_2001}. In our model, we find that $0<m<1$ at the transition so that it is indeed first-order.} In the limit of $T\to0$, we find that the break-point parameter $m \to 0$ at the transition, which suggests that the replica symmetry is restored at the $T=0$ quantum phase transition, 
similarly to Refs.~\cite{cugliandolo_quantum_2001,georges_mean-field_1999}. The phase diagram in the $(\Omega_0,T)$ plane for fixed values of $u$ and $v$ is shown in Fig.~\ref{fig:single_band_pd}. As can be seen in the figure, the region of the glass phase shrinks upon increasing $u$.

\subsection{Specific Heat}

Consider the specific heat $c$ of the disordered phase in the SB model. The temperature dependence of $c$ can be used as a thermodynamical diagnostic for crossover to the classical limit of the model. At high temperatures, we find that $c$ saturates to a constant value according to an anharmonic variant of the Dulong-Petit law (Eqn.~\ref{c_high_T}). Conversely, temperatures for which there is a large variation in the value of $c(T)$ are associated with a non-classical behavior. We will use this simple diagnostic to further illustrate that the dynamical Planckian regime of the model, discussed in Sec.~\ref{sec:DYNAMICS} (see also Fig.~\ref{fig:single_band_demo}), is indeed of quantum mechanical nature.

The computation of the specific heat is done as follows. We first use the disorder-averaged free-energy density to derive an expression for the internal energy $U = \partial (\beta \overline{f}) / \partial \beta$ of the disordered phase. Then, the specific heat $c = \partial U / \partial T$ can be evaluated numerically, or analytically at the high- and low-temperature limits (see Appendix~\ref{td_functions} for more details). At high temperatures, we find that 
\begin{equation}
    c \approx \frac{3}{4} + b\sqrt{\frac{\Omega_u}{T}}\quad,\quad T/\Omega_u \gg 1,
    \label{c_high_T}
\end{equation}
where $b = \left( (\Omega_0/\Omega_u)^2 - (\Omega_v/\Omega_u)^5/3 \right)/2$. The high-$T$ limit, where $c\to3/4$, is a result of the quartic term which becomes dominant as $T\gg \Omega_u$ (see further discussion in \ref{td_functions}).  At low temperatures the specific heat vanishes exponentially as the system is gapped at $T=0$. 

\begin{figure}[t]
\centering
\includegraphics[width=\columnwidth]{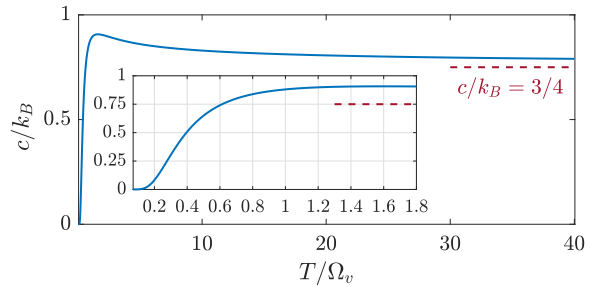}
\caption{Specific heat of the disordered phase of the SB model as a function of $T/\Omega_v$. The high-temperature classical limit $c = 3/4$ is denoted by the red dashed line. The inset shows a zoom-in on low to intermediate temperatures. Here, $\Omega_0/\Omega_v = 1.1$ and $u/\Omega_v^3 = 1.4$.} 
\label{fig:c_demo}
\end{figure}

In Fig.~\ref{fig:c_demo}, we show the $T$-dependence of $c$ for a representative set of parameters, which demonstrates that $c\to3/4$ at high temperatures, and vanishes as $T\to 0$ (see inset). We will further discuss the temperature dependence of $c$ and the correspondence between $c$ and $\tau_{\text{ph}}$ in Sec.~\ref{sec:DYNAMICS}.   

\section{Dynamics}
\label{sec:DYNAMICS}

In this section, we discuss the dynamical properties of the disordered phase of the SB model. We identify three dynamical regimes that posses different spectral properties for which the phonon lifetime $\tau_{\text{ph}}$ is a convenient probe. In particular, we demonstrate the existence of an intermediate-$T$ `Planckian' regime, where  $\tau_{\text{ph}}$ is of the order of the Planckian time scale $\tau_{\text{Pl}}=\hbar/T$.

Focusing on the disordered phase, where the model is self-averaging, we may use the framework of the Keldysh formalism (see Appendix \ref{sec:keldysh_formalism}) to compute the disordered-averaged partition function, which can be expressed as a functional integral of an effective Keldysh action. The functional integral is controlled by the saddle point of the effective Keldysh action in the limit of $N \to \infty$. This enables us to obtain a set of self-consistent equations for the well-known retarded, advanced and Keldysh Green's functions, from which one can extract the desired dynamical information. The Keldysh saddle-point equations are given by \begin{eqnarray}
 \hat{G}_{R}\left(\omega\right) &=& -\frac{1}{\omega^{2}-\Omega_{0}^{2}-\hat{\Pi}_{R}\left(\omega\right)} \nonumber , \\
 \hat{G}_{K}\left(\omega\right)&=&2i\coth\left(\frac{\beta\omega}{2}\right)\text{Im}\left[\hat{G}_{R}\left(\omega\right)\right], \nonumber   \\
\Pi_R\left(t\right) &=& iv^{2} G_{R}\left( t \right) G_{K}\left( t \right) -\frac{iu}{2}G_{K}\left(t\right)\delta(t) ,
 \label{eq:Keldysh_sc_eqs}
\end{eqnarray}
where the Keldysh component is set according to the fluctuation dissipation theorem (FDT) to enforce thermal equilibrium. 

For a wide range of temperatures, Eqs.~\ref{eq:Keldysh_sc_eqs} can be numerically solved by an iterative procedure (see Appendix \ref{sec:numerical_methods}). Solving these equations enables us to obtain the spectral function and the phonon lifetime $\tau_{\text{ph}}$, defined through the late time behavior of the retarded propagator: $G_R(t)\propto e^{-t/\tau_{\text{ph}}}$. 

Fig.~\hyperref[fig:single_band_demo]{\ref{fig:single_band_demo}a,b} shows the dependence of $\tau_{\text{ph}}$, normalized by $\tau_{\text{Pl}}=\hbar/T$, on $T$ for a system with $\Omega_0 \sim \Omega_v \sim \Omega_u$. The parameters are chosen such that the system is always far from the glass phase. 
We identify three distinct temperature regimes.
At low temperatures, $T\ll \Omega_0$, we find that $\tau_{\text{ph}}$ rises sharply with decreasing $T$; this behavior is associated with the finite gap at $T=0$. At high temperatures, $T\gg \Omega_0$, $\tau_{\text{ph}}$ becomes temperature independent, and hence $T \tau_{\text{ph}}/\hbar \gg 1$. As we will argue below, this is the classical regime, where $\tau_{\text{ph}} \propto \sqrt{M}$ and is independent of $\hbar$. 
$T \tau_{\text{ph}}/\hbar$ has a shallow minimum at intermediate temperatures. 



We refer to this intermediate regime as the `Planckian regime', where $\tau_{\text{ph}} \sim \hbar/T$. 
Note that within our model, this regime is not parametrically large, but rather extends over a finite range of temperatures around $T\sim \Omega_0$. The minimal value of $T \tau_{\text{ph}}/\hbar$ depends on the parameters $\Omega_0/\Omega_v$ and $\Omega_u/\Omega_v$. As we will show below (Fig.~\ref{fig:minimal_lifetime}), for an appropriate choice of these parameters (in the vicinity of the glass phase), we find that $T \tau_{\text{ph}}/\hbar$ approaches 1. 


At the two limiting cases of high and low temperatures, Eqs.~\ref{eq:Keldysh_sc_eqs} are amenable to analytical approximations, from which one can extract the qualitative behavior of the phonon lifetime. 
To carry out these approximations, we assume that (a) the phonons are well-defined quasiparticles at these regimes, and that (b) there are only two energy scales in the system, corresponding to the renormalized phonon frequency and the phonon lifetime. We then use these assumptions to make the following ansatz for the retarded Green's function,
\begin{equation}
    -\hat{G}_R^{-1}(\omega) = \omega^2 - \Pi_0 + 2i\omega/\tau_{\text{ph}},
    \label{retarded_ansatz}
\end{equation}
where (a) means that we assume $\sqrt{\Pi_0} \gg \tau_{\text{ph}}^{-1}$ and we use (b) to ignore higher order terms in the retarded self-energy. Note that the ansatz in Eqn.~\ref{retarded_ansatz} contains two unknown quantities: $\tau_{\text{ph}}$ and $\Pi_0$. 
Note also that $\Pi_0$ is a thermodynamic quantity, that corresponds to the zero-frequency phonon stiffness. Details on the imaginary- and real-time derivations for $\Pi_0$ are found in Appendices \ref{sec:instability} and \ref{sec:real_time_Pi_0}, respectively, and in Appendix \ref{sec:estimate_tau} we give details on the estimation of $\tau_\text{ph}$.


At low temperatures ($T\ll \Omega_0$), the system is essentially gapped. 
We then expect the phonons to be exponentially long-lived, since scattering off thermal excitations is exponentially rare. We confirm this expectation and find that 
\begin{equation}
    \tau_\text{ph} \sim \frac{\sqrt{\Pi_0T} }{v}e^{\sqrt{\Pi_0}/2T}.
    \label{eq:tau_low_T}
\end{equation}

At high temperatures ($T\gg \Omega_u$), we find that $\Pi_0 \approx \sqrt{uT}$ and 
\begin{equation}
    \tau_\text{ph} \sim \frac{\Omega_v^{-1}}{r^{3/2}}, 
    \label{eq:tau_high_T}
\end{equation}
where $r=\Omega_v/\Omega_u = v^{2/5}/u^{1/3}$ is a dimensionless number of order one ($r\leq 1$ in our setting of interest). 
Interestingly, it appears that the phonon lifetime becomes independent of temperature, which agrees with the numerical data shown in  Fig.~\hyperref[fig:single_band_demo]{\ref{fig:single_band_demo}b}. Moreover, by reinstating $\hbar$ and $M$ in Eqn.~\ref{eq:tau_high_T} (using the definitions of $\Omega_u$ and $\Omega_v$ above Eqn.~\ref{eq:L}), we find that 
\begin{equation}
    \tau_\text{ph} \propto \sqrt{M}
    \label{classical_lifetime_prop_to_sqrtM}
\end{equation}
and is independent of $\hbar$. This suggests that for $T\gg \Omega_u$, the model obeys classical dynamics (see also \ref{sec:high_T_lifetime}). Specifically, the dynamics of underdamped harmonic oscillators, as $\sqrt{\Pi_0} \gg 1/\tau_\text{ph}$. To see that the aforementioned inequality holds, we use the fact that at high temperatures $\Pi_0 = \sqrt{uT}$ (see Eqn.~\ref{eq:pos_Pi_ex}). Then, substituting $\tau_\text{ph}$ and $\Pi_0$, we see that the inequality holds only if $T \gg r^{10} \Omega_u$. This is indeed the case, because $r\leq 1$ and $T\gg \Omega_u$.

Note that the parameters in Fig.~\hyperref[fig:single_band_demo]{\ref{fig:single_band_demo}a,b} (with $\Omega_0/\Omega_v = 1.1$) are identical to the ones in Fig.~\ref{fig:c_demo}.
It is then interesting to examine the correspondence between the behavior of the specific heat and the phonon lifetime. At low temperatures, we find that $c$ and $1/\tau_\text{ph}$ vanishes exponentially, as expected due to fact that the system is gapped at $T=0$. At high temperatures, the system approaches the classical limit, as can be seen thermodynamically by the fact that $c$ approaches a constant value, and dynamically as $\tau_\text{ph}\propto \sqrt{M}$. Importantly, at intermediate temperatures ($T/\Omega_v \approx 0.45$ to $T/\Omega_v \approx 1.15$), which we referred to as the Planckian regime in Fig.~\hyperref[fig:single_band_demo]{\ref{fig:single_band_demo}a}, we find a significant variation in the value of $c$, which serves as another indication of the quantum mechanical nature of this dynamical regime.

\subsubsection{Minimal phonon lifetime}

\begin{figure}[H]
    \centering
    \includegraphics[width=\columnwidth]{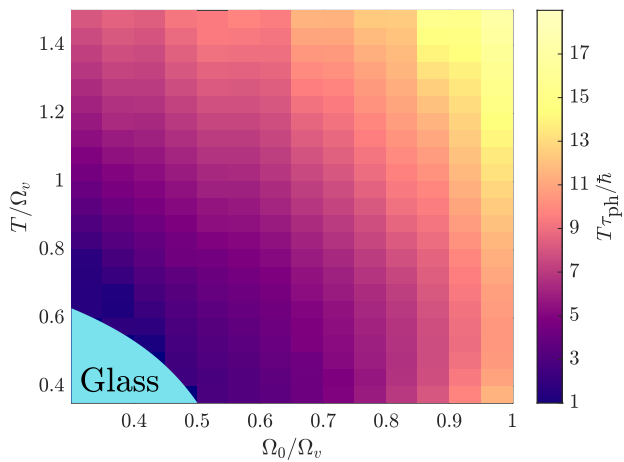}
    \caption{Phonon lifetime in units of $\tau_\text{Pl}=\hbar/T$ in the disordered phase of the SB model as a function of $\Omega_0/\Omega_v$ and $T/\Omega_v$. The bottom left corner contains a region which realizes the glass phase (cyan area). Observe that the phonon lifetime is approaching the Planckian time scale at the vicinity of the glass phase. The minimal coefficient in the figure is $\alpha \approx 1.01$ for $\Omega_0 / \Omega_v=0.35$ and $T/ \Omega_v = 0.55$. Here, $u/\Omega_v^3=1.8$.}
\label{fig:minimal_lifetime}
\end{figure}

We have thus far demonstrated that the model has an intermediate-temperature dynamical regime where the phonon lifetime is of the order of the Planckian time scale
\begin{equation}
       \tau_\text{ph} = \alpha \frac{\hbar}{k_B T},
       \label{eq:Planckian_coeff}
\end{equation}
with a numerical coefficient $\alpha$. It is interesting to ask what is the minimal attainble value of $\alpha$ within our model. 
For a generic choice of parameters in the strongly-coupled regime, where
\begin{equation}
    \Omega_0 \sim \Omega_v \sim \Omega_u,
    \label{strong_coupling_regime}
\end{equation} 
one finds that the numerical coefficient $\alpha$ in Eqn.~\ref{eq:Planckian_coeff} is of the order of $10$ around the Planckian regime, as demonstrated in Fig.~\ref{fig:single_band_demo}. 
However, as one approaches the vicinity of the glass phase in the $(\Omega_0,T)$ plane (for fixed $u$ and $v$), this numerical coefficient tends to decrease. In particular, we find that for sufficiently large values of $u$, which enable us to approach relatively small values of $\Omega_0$ and $T$ and remain in the disordered phase, $\alpha$ reaches values close to unity, see Fig.~\ref{fig:minimal_lifetime}.  

Scanning the $(\Omega_0,T)$ parameter space, we find 
that $ \alpha \gtrsim 1$, and that $\alpha \approx 1$ for regions with $\Omega_0 < \Omega_c(u,v)$ and temperatures slightly above the glass transition, where $ \Omega_c(u,v)$ is the $T=0$ glass transition frequency. This is apparent in Fig.~\ref{fig:minimal_lifetime}. The picture for other values of $u$ is similar. Interestingly, we find that $\alpha$ never drops below $1$, supporting the conjecture of a universal bound on $\alpha$.

\section{Generalization to multiple phonon branches}
\label{sec:MB_model}
Considering the simple, SB version of the model enabled us to obtain its phase diagram and specific heat, and served as a convenient platform for the study of its real-time dynamics. This version, however, describes a rather artificial setting in terms of phonons, where we consider $N$ degenerate optical phonon branches (with $\Omega_i=\Omega_0$ for all $i$'s). In physical insulating compounds, optical phonon branches are typically spread over a finite bandwidth, rather than being degenerate. 
As a step towards making our model more realistic, we consider a multi-phonon branch generalization, where the $\Omega_i$'s satisfy $\Omega_1 \leq ... \leq \Omega_N$. We dub this version the `multi-branch' (MB) model. The large-$N$ limit is taken such that the distribution of frequencies obeys $\rho(\Omega) \equiv \sum_{i=1}^N \delta(\Omega - \Omega_i) \rightarrow N f(\Omega)$, where $f(\Omega)$ is a function normalized such that $\int d\Omega f(\Omega) = 1$. The support of $f(\Omega)$ extends from $\Omega_{\text{min}}$ to $\Omega_{\text{max}}$, where $\Omega_{\text{max}}-\Omega_{\text{min}}$ is the bandwidth of the model. The SB model is recovered for $f(\Omega) = \delta(\Omega-\Omega_0)$. 

We consider the Green's function for the $i$th branch, defined by $G_i\left(\tau\right)\equiv \left<\phi_i\left(\tau\right)\phi_i\left(0\right)\right>$ (no sum). The generalization of the imaginary-time self-consistent equations in the disordered (replica-diagonal) phase is given by
\begin{eqnarray}
    \hat{G}_i\left(i\omega_{n}\right) &=& \frac{1}{ \omega_{n}^{2} + \Omega_{i}^2   - \hat{\Pi}\left(i\omega_{n}\right)}
, 
\\
\Pi\left(\tau\right) &=& \frac{v^{2}}{N^2}\sum_{i,j} G_i\left(\tau\right)G_j\left(\tau\right)-\frac{u}{N}\sum_i G_i\left(\tau\right)\delta\left(\tau\right). \nonumber \\
\label{eq:G_Pi_diagonal_many_bands}  \nonumber
\end{eqnarray}

As in the SB case, we first need to determine the phase diagram of the model. 
The replica analysis for the MB model is more complicated than in the SB case. Instead of calculating the phase diagram explicitly, we use a simple argument to bound the glass phase in the $(\Omega_{\text{min}},T)$ plane. Consider deforming the mode distribution function $f(\Omega)$ continuously to that of a SB model with $f_\text{SB}(\Omega) = \delta(\Omega - \Omega_{\text{min}})$. Such a deformation softens the phonon modes, stabilizing configurations with large equilibrium displacements. We therefore expect that the deformation expands the regime of the glass phase. Indeed, in the SB model, decreasing $\Omega_0$ brings us closer to the glass regime (Fig.~\ref{fig:single_band_pd}). 
Therefore, we assume that for fixed $\Omega_{v,u}$ and   $\Omega_{\text{min}}>\Omega_c(u,v)$ (where $\Omega_c$ is the location of the $T = 0$ glass transition in the corresponding SB model), the MB model is in the disordered phase.


\begin{figure}[t]
    \centering
    \includegraphics[width=\columnwidth]{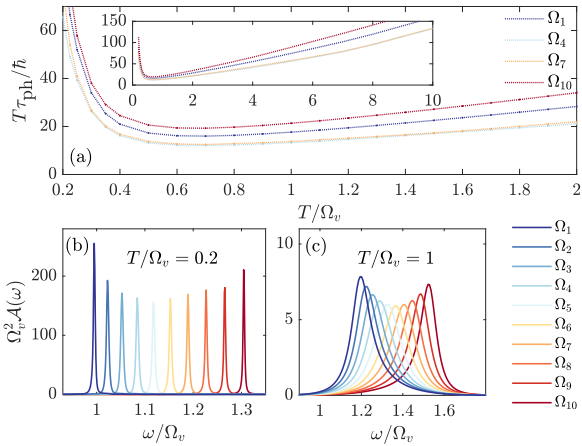}
    \caption{Dynamical properties of the disordered phase of the MB model. (a) shows the phonon lifetime in the units of the Planckian time-scale $\hbar/T$ as a function of $T$ for 4 out of 10 distinct modes. (b) and (c) presents the spectral function for relatively low and intermediate temperatures, respectively. Here, $\Omega_l,\text{ }l=1,...,10$, are uniformly distributed from $\Omega_\text{min}/\Omega_v = 1/2$ to $\Omega_\text{max}/\Omega_v = 1$; the $T\to0$ quantum phase transition in the corresponding SB model with $\Omega_0/\Omega_v=0.5$ occurs at $\Omega_c/\Omega_v \approx 0.41$; and $u/\Omega_v^3 = 1.9$.}
\label{fig:many_bands_demo}
\end{figure}

We now consider the real-time dynamics of disordered phase of the MB model. Here, the $i$th phonon branch is characterized by its corresponding spectral function $\mathcal{A}^{(i)} \left(\omega \right)$ and phonon lifetime $\tau_{\text{ph}}^{(i)}$. These are obtained by solving the generalized Keldysh saddle-point equations, which are given by Eqs.~\ref{eq:Keldysh_many_bands_sc_eqs} in Appendix \ref{sec:keldysh_formalism}. Fig.~\ref{fig:many_bands_demo} shows the dynamical properties for a representative set of parameters of the MB model\footnote{The numerical solution of Eqs.~\ref{eq:G_Pi_diagonal_many_bands} (and their corresponding real-time Keldysh Eqs.~\ref{eq:Keldysh_many_bands_sc_eqs}) is done by setting $f(\Omega) = \frac{1}{L}\sum_{l=1}^{L}\delta(\Omega - \Omega_l)$ where $\Omega_l$ are uniformly distributed between $\Omega_\text{min}$ and $\Omega_\text{max}$. This means that we are dividing the $\Omega_i$'s into $L$ subsets of size $N/L$. Consequently, sums of the form $(1/N)\sum_{i=1}^N G_i$ transform into $(1/L)\sum_{l=1}^L G_l$. \label{foot_note_implementation} }.

In Fig.~\hyperref[fig:many_bands_demo]{\ref{fig:many_bands_demo}a} we show the dependence of $\tau_{\text{ph}}^{(i)}/\tau_{\text{Pl}}$ on $T$ for $4$ out of $10$ distinct branches. Note that two of these branches are at the edges of the spectrum, $\Omega_1$ and $\Omega_{10}$, and two are near its center, $\Omega_4$ and $\Omega_7$. Here, the parameters are chosen such that the system is always far from the glass phase, by the assumption stated above. We find that the overall trends are similar to those found in the SB model (Fig.~\hyperref[fig:single_band_demo]{\ref{fig:single_band_demo}a}). In particular, we identify the same three dynamical regimes: the low-$T$ semiclassical regime where the lifetime diverges exponentially, the high-$T$ classical regime where the $\tau^{(i)}_{\text{ph}}$'s are independent of $T$, and an intermediate regime where $\tau^{(i)}_{\text{ph}}\sim \tau_{\text{Pl}}$. Interestingly, we observe that the minimum of $T \tau^{(i)}_{\text{ph}}/\hbar$ is flatter in the MB model compared to an SB model with roughly the same parameters. 
This allows for an expanded ``Planckian'' regime in the MB model. Moreover, we observe that the lifetimes of the individual modes are different: they are minimal near the center of the spectrum, and grow as the frequency approaches the top and bottom of the spectrum. 

In Fig.~\hyperref[fig:many_bands_demo]{\ref{fig:many_bands_demo}b,c} we show the spectral function of the 10 distinct branches for two temperatures, where the significant broadening of the spectral peaks demonstrates the crossover from the semiclassical regime, with $T/\Omega_v = 0.2$, to the Planckian regime of the MB model, with $T/\Omega_v = 1$. 

\section{Relation to the SYK Model}

It is natural to ask whether our model realizes a critical point with an emergent conformal symmetry at low energies, similarly to the SYK model \cite{Kitaev_SYK_talk,maldacena_remarks_2016}. 
Unfortunately, the answer appears to be no, both in our model, and also in general bosonic variants of the SYK model, as was recently discussed in \cite{baldwin_quenched_2019}.\footnote{See Ref.~\cite{facoetti_classical_2019} for an interesting discussion on the relation between classical glassy dynamics and the SYK model.} 

We begin by showing why the most naive approach to finding a conformally invariant point is inconsistent, and then we will show that the consistent approach leads to an unstable conformally invariant critical point, which is not the physical solution of the saddle-point equations. 
Naively, the first step towards an emegernt conformal (time-reparametrization) symmetry at low energies is to bring the self-consistent equations to a reparametrization-invariant form. Observing Eqs.~\ref{eq:G_Pi_diagonal} at $T=0$, this can be done by neglecting the $\omega^2$ term, while tuning $\Omega_0$ and $u$ to zero\footnote{Setting $u=0$ implies that the model is not well-defined at the full nonperturbative level. However, it can still be formally considered at the $N\to\infty$ limit for low temperatures, as long as $ v^{2/5} \lesssim \Omega_0$, see discussion below Eqn.~\ref{eq:low_T_Pi_0} in Appendix \ref{sec:replica_formalism}.}. This leads to the following set of `SYK-like' self-consistent equations:
\begin{eqnarray}
    \hat{G}\left(i\omega\right) &=& -\frac{1}{\hat{\Pi}\left(i\omega\right)}
\nonumber, 
\\
\Pi\left(\tau\right) &=& v^{2}G\left(\tau\right)^2
\label{eq:G_Pi_diagonal_SYK}.
\end{eqnarray}

Then, following the analogy with the SYK model, we substitute a scaling ansatz, given by $G_\text{conf}\left(\tau\right)\equiv b \left|\tau\right|^{-2\Delta_\phi}$, into the self-consistent equations, and obtain the scaling dimension $\Delta_\phi$ and the numerical coefficient $b$. However, the solution obtained by this approach is an inconsistent solution of Eqs.~\ref{eq:G_Pi_diagonal_SYK}. The inconsistency comes from the UV behavior of the self-energy. Namely, the UV piece of the self-energy is non-negligible:
\begin{eqnarray}
    \hat{\Pi}\left(i\omega\right) &=& \Pi_{\text{UV}} + \hat{\Pi}_\text{conf}(i\omega),
\end{eqnarray}
where $\hat{\Pi}_\text{conf}(i\omega) \equiv - \hat{G}^{-1}_\text{conf}(i\omega)$. That is, the leading term of $\Pi(i\omega \to 0)$ is a constant, $\Pi_{\text{UV}}$, rather than the conformal self-energy $\hat{\Pi}_\text{conf}(i\omega)$ with which we started our consistency check. In addition, due to its positive sign, this constant violates the positivity condition that $\hat{G}(i\omega) > 0$, since $\hat{G}(i\omega=0) = -1/\Pi_{\text{UV}} < 0$. This behavior originates from the bosonic nature of the degrees of freedom, which manifests itself in the even parity of the Green's function. In the fermionic case, the odd parity of the Green's function enables one to neglect the UV piece of the Green's function safely. 

This inconsistency can be fixed by reinstating $\Omega_0$, and using it as a counter-term, such that the self-energy reads
\begin{equation}
    \hat{\Pi}(i\omega) = \Pi_{\text{UV}} + \hat{\Pi}_\text{conf}(i\omega) - \Omega_0^2.
\end{equation}
We can then fine tune $\Omega_0$ to cancel the non-conformal constant by letting $\Omega_{0}^2\to\Omega_\text{conf}^2 \equiv \Pi_{\text{UV}}$, giving
\begin{equation}
    \hat{\Pi}(i\omega) = \hat{\Pi}_\text{conf}(i\omega).
\end{equation}
The scaling ansatz is then a consistent solution of the saddle-point equations at low temperatures and long times, which can be verified both analytically and numerically (see \ref{relation_to_SYK} and \ref{imag_time_for_dis_phase}, and also Section 2 of \cite{chang_melonic_2018}). 

Interestingly, there are several warning signs indicating that this scale-invariant solution is not realized in our model. The most obvious one in our setting is the fact that for $\Omega_0=\Omega_\text{conf}$ and $u\to 0$, the system realizes the glass phase associated with the 1SRSB solution of the saddle-point equations. However, even within the disordered phase, it turns out that the conformal solution is unstable.

In Ref.~\cite{Giombi2017}, similar conformally-invariant self-consistent equations were studied in the context of bosonic tensor models (without disorder), with $q$-body rather than $3$-body interactions ($q\geq 4)$. 
Notably, \cite{Giombi2017} found that the scaling dimension of the $\phi^2$ composite operator is complex for the conformal field theory (CFT) associated with this form of self-consistent equations. This violates the unitarity condition and implies that the CFT is an unstable solution of the theory. This is also true in our model \cite{Kleb_private}, and it suggests that there exists another, stable solution of the disordered saddle-point equations with $\Omega_0=\Omega_\text{conf}$. Indeed, we find that such a solution exists. This solution is gapped at $T=0$, and by comparing the free energies of the two solutions, we also find that it is thermodynamically favorable (see further discussion below Eqn.~\ref{eq:modified_updating_scheme} in Appendix \ref{sec:numerical_methods}).  A similar observation was made in \cite{azeyanagi_phase_2018}.  

\label{relation_to_syk_main_text}


\section{Discussion and Outlook}
\label{disc}
 
 In this work, we have studied a solvable model of $N$ interacting phonons. In the limit $N\rightarrow \infty$, the model is solvable for any interaction strength and temperature, allowing us to access the strongly interacting regime. At low temperature and strong interactions, the system undergoes a first order transition into a replica symmetry breaking (glass) phase. Focusing on the dynamics in the replica diagonal (disordered) phase, we find that the system crosses over between three distinct regimes as the temperature increases: a semiclassical regime with long-lived quasiparticle (phonon) excitations at low temperatures, a classical regime at high temperatures, and an intermediate strongly-interacting ``phonon fluid'' regime. In the latter regime, the minimal phonon relaxation time is of the order of the Planckian time scale, $\tau_\text{ph} = \alpha \hbar/T$, with $\alpha$ approaching unity near the transition to the glass phase. 
 
 Our work was motivated by measurements of the thermal diffusivity in a broad class of insulating materials, indicating that these systems may indeed be described as a strongly coupled liquid of phonons, with a relaxation time that approaches the Planckian time. Clearly, our model is not meant to realistically model any material; rather, it provides a concrete example of such a strongly interacting quantum regime in a bosonic system. Within our model, this regime is realized over an intermediate temperature range; at sufficiently high temperatures the system always crosses over to a classical regime, at which $\tau_\text{ph} \gg \hbar/T$ and the specific heat approaches its classical limit. Interestingly, we find that in the intermediate quantum regime, the relaxation time never drops below $\hbar/T$, consistent with the notion of a universal ``Planckian bound'' on thermalization times.   
 
From a theoretical perspective, our model can be viewed as a bosonic variant of the SYK model. However, there are crucial differences between our model and the fermionic SYK model. In particular, we showed that the emergent low-energy conformally invariant saddle-point solution of this model is not realized, as it is found deep inside the glass phase. This is in line with general arguments regarding the low-temperature behavior of bosonic SYK-like models~\cite{baldwin_quenched_2019} and the presence of operators with complex scaling dimensions at the putative conformally invariant point~\cite{Giombi2017}.

{Some natural questions remain open. It is interesting to study the correspondence between the phonon inverse lifetime and Lyapunov exponent which characterizes the growth of out-of-time-order correlation functions~\cite{Larkin1969,Kitaev_SYK_talk}. At the high- and low-temperature limits, where the phonon lifetime is very long, one might expect similar qualitative behavior~\cite{grozdanov_kinetic_2019}. It is not clear, however, if this expected qualitative correspondence will extend to the ``phonon fluid'' regime. Especially in light of the fact that no such correspondence was found in similar models \cite{Mao_Chowdhury_Senthil_2019,cheng_chaos_2019}.}     

Furthermore, to address the transport properties in the quantum phonon fluid regime, our model needs to be generalized to higher dimensions. This can be done, e.g., by placing a copy of our model on each site of a $D-$dimensional lattice, along the lines of Ref.~\cite{Balents}. Related to this issue is the absence of acoustic phonon modes in our model. These are protected by Goldstone's theorem, and must remain gapless and long-lived even in the presence of strong interactions. Nevertheless, in the strongly coupled ``phonon fluid'' regime, their contribution to transport may be negligible due to their small phase space. 

Another natural question regards the effect of glassiness on the dynamics in our model.
In the glass phase, ergodicity is violated and the phase space is fragmented into disconnected clusters. However, intuitively, one may expect the relaxation dynamics within each phase space cluster to be qualitatively similar to that of the disordered phase. We leave a detailed investigation of this question to future studies.

 
\acknowledgements 
We thank E. Altman, D. Arovas, D. Chowdhury, L. Cugliandolo, A. Kapitulnik, S. Kivelson, I. Klebanov, S. Sachdev, T. Senthil for useful discussions throughout this work. 
 EB was supported by the European Research Council (ERC) under grant HQMAT (grant no. 817799) and by the US-Israel Binational Science Foundation (BSF).

\bibliography{refs,nfl}

\begin{thebibliography}{52}
\expandafter\ifx\csname natexlab\endcsname\relax\def\natexlab#1{#1}\fi
\expandafter\ifx\csname bibnamefont\endcsname\relax
  \def\bibnamefont#1{#1}\fi
\expandafter\ifx\csname bibfnamefont\endcsname\relax
  \def\bibfnamefont#1{#1}\fi
\expandafter\ifx\csname citenamefont\endcsname\relax
  \def\citenamefont#1{#1}\fi
\expandafter\ifx\csname url\endcsname\relax
  \def\url#1{\texttt{#1}}\fi
\expandafter\ifx\csname urlprefix\endcsname\relax\def\urlprefix{URL }\fi
\providecommand{\bibinfo}[2]{#2}
\providecommand{\eprint}[2][]{\url{#2}}

\bibitem[{\citenamefont{Zaanen}(2004)}]{zaanen_why_2004}
\bibinfo{author}{\bibfnamefont{J.}~\bibnamefont{Zaanen}},
  \bibinfo{journal}{Nature} \textbf{\bibinfo{volume}{430}},
  \bibinfo{pages}{512} (\bibinfo{year}{2004}), ISSN \bibinfo{issn}{0028-0836,
  1476-4687}, \urlprefix\url{http://www.nature.com/articles/430512a}.

\bibitem[{\citenamefont{Sachdev}(2011)}]{sachdev_quantum_2011}
\bibinfo{author}{\bibfnamefont{S.}~\bibnamefont{Sachdev}},
  \emph{\bibinfo{title}{Quantum Phase Transitions}}
  (\bibinfo{publisher}{Cambridge University Press}, \bibinfo{year}{2011}),
  \bibinfo{edition}{2nd} ed.

\bibitem[{\citenamefont{Bruin et~al.}(2013)\citenamefont{Bruin, Sakai, Perry,
  and Mackenzie}}]{bruin_similarity_2013}
\bibinfo{author}{\bibfnamefont{J.~a.~N.} \bibnamefont{Bruin}},
  \bibinfo{author}{\bibfnamefont{H.}~\bibnamefont{Sakai}},
  \bibinfo{author}{\bibfnamefont{R.~S.} \bibnamefont{Perry}}, \bibnamefont{and}
  \bibinfo{author}{\bibfnamefont{A.~P.} \bibnamefont{Mackenzie}},
  \bibinfo{journal}{Science} \textbf{\bibinfo{volume}{339}},
  \bibinfo{pages}{804} (\bibinfo{year}{2013}), ISSN \bibinfo{issn}{0036-8075,
  1095-9203},
  \urlprefix\url{https://science.sciencemag.org/content/339/6121/804}.

\bibitem[{\citenamefont{Legros et~al.}(2019)\citenamefont{Legros, Benhabib,
  Tabis, Laliberté, Dion, Lizaire, Vignolle, Vignolles, Raffy, Li
  et~al.}}]{legros_universal_2019}
\bibinfo{author}{\bibfnamefont{A.}~\bibnamefont{Legros}},
  \bibinfo{author}{\bibfnamefont{S.}~\bibnamefont{Benhabib}},
  \bibinfo{author}{\bibfnamefont{W.}~\bibnamefont{Tabis}},
  \bibinfo{author}{\bibfnamefont{F.}~\bibnamefont{Laliberté}},
  \bibinfo{author}{\bibfnamefont{M.}~\bibnamefont{Dion}},
  \bibinfo{author}{\bibfnamefont{M.}~\bibnamefont{Lizaire}},
  \bibinfo{author}{\bibfnamefont{B.}~\bibnamefont{Vignolle}},
  \bibinfo{author}{\bibfnamefont{D.}~\bibnamefont{Vignolles}},
  \bibinfo{author}{\bibfnamefont{H.}~\bibnamefont{Raffy}},
  \bibinfo{author}{\bibfnamefont{Z.~Z.} \bibnamefont{Li}},
  \bibnamefont{et~al.}, \bibinfo{journal}{Nature Physics}
  \textbf{\bibinfo{volume}{15}}, \bibinfo{pages}{142} (\bibinfo{year}{2019}),
  ISSN \bibinfo{issn}{1745-2481},
  \urlprefix\url{https://www.nature.com/articles/s41567-018-0334-2}.

\bibitem[{\citenamefont{Cao et~al.}(2020)\citenamefont{Cao, Chowdhury,
  Rodan-Legrain, Rubies-Bigorda, Watanabe, Taniguchi, Senthil, and
  Jarillo-Herrero}}]{cao_strange_2020}
\bibinfo{author}{\bibfnamefont{Y.}~\bibnamefont{Cao}},
  \bibinfo{author}{\bibfnamefont{D.}~\bibnamefont{Chowdhury}},
  \bibinfo{author}{\bibfnamefont{D.}~\bibnamefont{Rodan-Legrain}},
  \bibinfo{author}{\bibfnamefont{O.}~\bibnamefont{Rubies-Bigorda}},
  \bibinfo{author}{\bibfnamefont{K.}~\bibnamefont{Watanabe}},
  \bibinfo{author}{\bibfnamefont{T.}~\bibnamefont{Taniguchi}},
  \bibinfo{author}{\bibfnamefont{T.}~\bibnamefont{Senthil}}, \bibnamefont{and}
  \bibinfo{author}{\bibfnamefont{P.}~\bibnamefont{Jarillo-Herrero}},
  \bibinfo{journal}{Physical Review Letters} \textbf{\bibinfo{volume}{124}},
  \bibinfo{pages}{076801} (\bibinfo{year}{2020}), \bibinfo{note}{publisher:
  American Physical Society},
  \urlprefix\url{https://link.aps.org/doi/10.1103/PhysRevLett.124.076801}.

\bibitem[{\citenamefont{Zhang et~al.}(2017)\citenamefont{Zhang, Levenson-Falk,
  Ramshaw, Bonn, Liang, Hardy, Hartnoll, and
  Kapitulnik}}]{zhang_anomalous_2017}
\bibinfo{author}{\bibfnamefont{J.}~\bibnamefont{Zhang}},
  \bibinfo{author}{\bibfnamefont{E.~M.} \bibnamefont{Levenson-Falk}},
  \bibinfo{author}{\bibfnamefont{B.~J.} \bibnamefont{Ramshaw}},
  \bibinfo{author}{\bibfnamefont{D.~A.} \bibnamefont{Bonn}},
  \bibinfo{author}{\bibfnamefont{R.}~\bibnamefont{Liang}},
  \bibinfo{author}{\bibfnamefont{W.~N.} \bibnamefont{Hardy}},
  \bibinfo{author}{\bibfnamefont{S.~A.} \bibnamefont{Hartnoll}},
  \bibnamefont{and}
  \bibinfo{author}{\bibfnamefont{A.}~\bibnamefont{Kapitulnik}},
  \bibinfo{journal}{Proceedings of the National Academy of Sciences}
  \textbf{\bibinfo{volume}{114}}, \bibinfo{pages}{5378} (\bibinfo{year}{2017}),
  ISSN \bibinfo{issn}{0027-8424, 1091-6490},
  \urlprefix\url{https://www.pnas.org/content/114/21/5378}.

\bibitem[{\citenamefont{Davison et~al.}(2014)\citenamefont{Davison, Schalm, and
  Zaanen}}]{davison_holographic_2014}
\bibinfo{author}{\bibfnamefont{R.~A.} \bibnamefont{Davison}},
  \bibinfo{author}{\bibfnamefont{K.}~\bibnamefont{Schalm}}, \bibnamefont{and}
  \bibinfo{author}{\bibfnamefont{J.}~\bibnamefont{Zaanen}},
  \bibinfo{journal}{Physical Review B} \textbf{\bibinfo{volume}{89}},
  \bibinfo{pages}{245116} (\bibinfo{year}{2014}), ISSN
  \bibinfo{issn}{1098-0121, 1550-235X},
  \urlprefix\url{https://link.aps.org/doi/10.1103/PhysRevB.89.245116}.

\bibitem[{\citenamefont{Blake et~al.}(2018)\citenamefont{Blake, Lee, and
  Liu}}]{blake_quantum_2018}
\bibinfo{author}{\bibfnamefont{M.}~\bibnamefont{Blake}},
  \bibinfo{author}{\bibfnamefont{H.}~\bibnamefont{Lee}}, \bibnamefont{and}
  \bibinfo{author}{\bibfnamefont{H.}~\bibnamefont{Liu}},
  \bibinfo{journal}{Journal of High Energy Physics}
  \textbf{\bibinfo{volume}{2018}} (\bibinfo{year}{2018}), ISSN
  \bibinfo{issn}{1029-8479}, \bibinfo{note}{arXiv: 1801.00010},
  \urlprefix\url{http://arxiv.org/abs/1801.00010}.

\bibitem[{\citenamefont{Patel et~al.}(2017)\citenamefont{Patel, Chowdhury,
  Sachdev, and Swingle}}]{patel_quantum_2017}
\bibinfo{author}{\bibfnamefont{A.~A.} \bibnamefont{Patel}},
  \bibinfo{author}{\bibfnamefont{D.}~\bibnamefont{Chowdhury}},
  \bibinfo{author}{\bibfnamefont{S.}~\bibnamefont{Sachdev}}, \bibnamefont{and}
  \bibinfo{author}{\bibfnamefont{B.}~\bibnamefont{Swingle}},
  \bibinfo{journal}{Physical Review X} \textbf{\bibinfo{volume}{7}}
  (\bibinfo{year}{2017}), ISSN \bibinfo{issn}{2160-3308}, \bibinfo{note}{arXiv:
  1703.07353}, \urlprefix\url{http://arxiv.org/abs/1703.07353}.

\bibitem[{\citenamefont{Patel et~al.}(2018)\citenamefont{Patel, McGreevy,
  Arovas, and Sachdev}}]{patel_magnetotransport_2018}
\bibinfo{author}{\bibfnamefont{A.~A.} \bibnamefont{Patel}},
  \bibinfo{author}{\bibfnamefont{J.}~\bibnamefont{McGreevy}},
  \bibinfo{author}{\bibfnamefont{D.~P.} \bibnamefont{Arovas}},
  \bibnamefont{and} \bibinfo{author}{\bibfnamefont{S.}~\bibnamefont{Sachdev}},
  \bibinfo{journal}{Physical Review X} \textbf{\bibinfo{volume}{8}}
  (\bibinfo{year}{2018}), ISSN \bibinfo{issn}{2160-3308},
  \urlprefix\url{https://link.aps.org/doi/10.1103/PhysRevX.8.021049}.

\bibitem[{\citenamefont{Davison et~al.}(2017)\citenamefont{Davison, Fu,
  Georges, Gu, Jensen, and Sachdev}}]{davison_thermoelectric_2017}
\bibinfo{author}{\bibfnamefont{R.~A.} \bibnamefont{Davison}},
  \bibinfo{author}{\bibfnamefont{W.}~\bibnamefont{Fu}},
  \bibinfo{author}{\bibfnamefont{A.}~\bibnamefont{Georges}},
  \bibinfo{author}{\bibfnamefont{Y.}~\bibnamefont{Gu}},
  \bibinfo{author}{\bibfnamefont{K.}~\bibnamefont{Jensen}}, \bibnamefont{and}
  \bibinfo{author}{\bibfnamefont{S.}~\bibnamefont{Sachdev}},
  \bibinfo{journal}{Physical Review B} \textbf{\bibinfo{volume}{95}}
  (\bibinfo{year}{2017}), ISSN \bibinfo{issn}{2469-9950, 2469-9969},
  \urlprefix\url{http://link.aps.org/doi/10.1103/PhysRevB.95.155131}.

\bibitem[{\citenamefont{Chowdhury et~al.}(2018)\citenamefont{Chowdhury, Werman,
  Berg, and Senthil}}]{chowdhury_translationally_2018}
\bibinfo{author}{\bibfnamefont{D.}~\bibnamefont{Chowdhury}},
  \bibinfo{author}{\bibfnamefont{Y.}~\bibnamefont{Werman}},
  \bibinfo{author}{\bibfnamefont{E.}~\bibnamefont{Berg}}, \bibnamefont{and}
  \bibinfo{author}{\bibfnamefont{T.}~\bibnamefont{Senthil}},
  \bibinfo{journal}{Physical Review X} \textbf{\bibinfo{volume}{8}},
  \bibinfo{pages}{031024} (\bibinfo{year}{2018}), ISSN
  \bibinfo{issn}{2160-3308}, \bibinfo{note}{arXiv: 1801.06178},
  \urlprefix\url{http://arxiv.org/abs/1801.06178}.

\bibitem[{\citenamefont{Hartnoll}(2015)}]{hartnoll_theory_2015}
\bibinfo{author}{\bibfnamefont{S.~A.} \bibnamefont{Hartnoll}},
  \bibinfo{journal}{Nature Physics} \textbf{\bibinfo{volume}{11}},
  \bibinfo{pages}{54} (\bibinfo{year}{2015}), ISSN \bibinfo{issn}{1745-2481},
  \urlprefix\url{https://www.nature.com/articles/nphys3174}.

\bibitem[{\citenamefont{Nussinov}(2020)}]{Nussinov_2020}
\bibinfo{author}{\bibfnamefont{Z.}~\bibnamefont{Nussinov}},
  \bibinfo{journal}{Nuclear Physics B} \textbf{\bibinfo{volume}{953}},
  \bibinfo{pages}{114948} (\bibinfo{year}{2020}), ISSN
  \bibinfo{issn}{0550-3213},
  \urlprefix\url{http://www.sciencedirect.com/science/article/pii/S0550321320300341}.

\bibitem[{\citenamefont{Kovtun et~al.}(2005)\citenamefont{Kovtun, Son, and
  Starinets}}]{kovtun_viscosity_2005}
\bibinfo{author}{\bibfnamefont{P.~K.} \bibnamefont{Kovtun}},
  \bibinfo{author}{\bibfnamefont{D.~T.} \bibnamefont{Son}}, \bibnamefont{and}
  \bibinfo{author}{\bibfnamefont{A.~O.} \bibnamefont{Starinets}},
  \textbf{\bibinfo{volume}{94}}, \bibinfo{pages}{111601}
  (\bibinfo{year}{2005}), \bibinfo{note}{publisher: American Physical Society},
  \urlprefix\url{https://link.aps.org/doi/10.1103/PhysRevLett.94.111601}.

\bibitem[{\citenamefont{Shenker and Stanford}(2014)}]{shenker_black_2014}
\bibinfo{author}{\bibfnamefont{S.~H.} \bibnamefont{Shenker}} \bibnamefont{and}
  \bibinfo{author}{\bibfnamefont{D.}~\bibnamefont{Stanford}},
  \bibinfo{journal}{Journal of High Energy Physics}
  \textbf{\bibinfo{volume}{2014}}, \bibinfo{pages}{67} (\bibinfo{year}{2014}),
  ISSN \bibinfo{issn}{1029-8479},
  \urlprefix\url{https://doi.org/10.1007/JHEP03(2014)067}.

\bibitem[{\citenamefont{Roberts et~al.}(2015)\citenamefont{Roberts, Stanford,
  and Susskind}}]{roberts_localized_2015}
\bibinfo{author}{\bibfnamefont{D.~A.} \bibnamefont{Roberts}},
  \bibinfo{author}{\bibfnamefont{D.}~\bibnamefont{Stanford}}, \bibnamefont{and}
  \bibinfo{author}{\bibfnamefont{L.}~\bibnamefont{Susskind}},
  \bibinfo{journal}{Journal of High Energy Physics}
  \textbf{\bibinfo{volume}{2015}}, \bibinfo{pages}{51} (\bibinfo{year}{2015}),
  ISSN \bibinfo{issn}{1029-8479},
  \urlprefix\url{https://doi.org/10.1007/JHEP03(2015)051}.

\bibitem[{\citenamefont{Blake}(2016)}]{blake_universal_2016}
\bibinfo{author}{\bibfnamefont{M.}~\bibnamefont{Blake}},
  \bibinfo{journal}{Physical Review Letters} \textbf{\bibinfo{volume}{117}},
  \bibinfo{pages}{091601} (\bibinfo{year}{2016}),
  \urlprefix\url{https://link.aps.org/doi/10.1103/PhysRevLett.117.091601}.

\bibitem[{\citenamefont{Hartnoll et~al.}(2018)\citenamefont{Hartnoll, Lucas,
  and Sachdev}}]{hartnoll_holographic_2018}
\bibinfo{author}{\bibfnamefont{S.~A.} \bibnamefont{Hartnoll}},
  \bibinfo{author}{\bibfnamefont{A.}~\bibnamefont{Lucas}}, \bibnamefont{and}
  \bibinfo{author}{\bibfnamefont{S.}~\bibnamefont{Sachdev}}
  (\bibinfo{year}{2018}), \bibinfo{note}{arXiv: 1612.07324},
  \urlprefix\url{http://arxiv.org/abs/1612.07324}.

\bibitem[{\citenamefont{Gu et~al.}(2017)\citenamefont{Gu, Qi, and
  Stanford}}]{Gu17}
\bibinfo{author}{\bibfnamefont{Y.}~\bibnamefont{Gu}},
  \bibinfo{author}{\bibfnamefont{X.-L.} \bibnamefont{Qi}}, \bibnamefont{and}
  \bibinfo{author}{\bibfnamefont{D.}~\bibnamefont{Stanford}},
  \bibinfo{journal}{Journal of High Energy Physics}
  \textbf{\bibinfo{volume}{2017}}, \bibinfo{pages}{125} (\bibinfo{year}{2017}),
  ISSN \bibinfo{issn}{1029-8479},
  \urlprefix\url{https://doi.org/10.1007/JHEP05(2017)125}.

\bibitem[{\citenamefont{Zaanen et~al.}(2015)\citenamefont{Zaanen, Liu, Sun, and
  Schalm}}]{zaanen_liu_sun_schalm_2015}
\bibinfo{author}{\bibfnamefont{J.}~\bibnamefont{Zaanen}},
  \bibinfo{author}{\bibfnamefont{Y.}~\bibnamefont{Liu}},
  \bibinfo{author}{\bibfnamefont{Y.-W.} \bibnamefont{Sun}}, \bibnamefont{and}
  \bibinfo{author}{\bibfnamefont{K.}~\bibnamefont{Schalm}},
  \emph{\bibinfo{title}{Holographic Duality in Condensed Matter Physics}}
  (\bibinfo{publisher}{Cambridge University Press}, \bibinfo{year}{2015}).

\bibitem[{\citenamefont{Geng}(2020)}]{geng_non-local_2020}
\bibinfo{author}{\bibfnamefont{H.}~\bibnamefont{Geng}} (\bibinfo{year}{2020}),
  \bibinfo{note}{arXiv: 2005.00021},
  \urlprefix\url{http://arxiv.org/abs/2005.00021}.

\bibitem[{\citenamefont{Behnia and Kapitulnik}(2019)}]{behnia_lower_2019}
\bibinfo{author}{\bibfnamefont{K.}~\bibnamefont{Behnia}} \bibnamefont{and}
  \bibinfo{author}{\bibfnamefont{A.}~\bibnamefont{Kapitulnik}},
  \bibinfo{journal}{arXiv:1905.03551 [cond-mat]}  (\bibinfo{year}{2019}),
  \bibinfo{note}{arXiv: 1905.03551},
  \urlprefix\url{http://arxiv.org/abs/1905.03551}.

\bibitem[{\citenamefont{Martelli et~al.}(2018)\citenamefont{Martelli, Jiménez,
  Continentino, Baggio-Saitovitch, and Behnia}}]{martelli_thermal_2018}
\bibinfo{author}{\bibfnamefont{V.}~\bibnamefont{Martelli}},
  \bibinfo{author}{\bibfnamefont{J.~L.} \bibnamefont{Jiménez}},
  \bibinfo{author}{\bibfnamefont{M.}~\bibnamefont{Continentino}},
  \bibinfo{author}{\bibfnamefont{E.}~\bibnamefont{Baggio-Saitovitch}},
  \bibnamefont{and} \bibinfo{author}{\bibfnamefont{K.}~\bibnamefont{Behnia}},
  \bibinfo{journal}{Physical Review Letters} \textbf{\bibinfo{volume}{120}},
  \bibinfo{pages}{125901} (\bibinfo{year}{2018}), ISSN
  \bibinfo{issn}{0031-9007, 1079-7114}, \bibinfo{note}{arXiv: 1802.05868},
  \urlprefix\url{http://arxiv.org/abs/1802.05868}.

\bibitem[{\citenamefont{Zhang et~al.}(2019)\citenamefont{Zhang, Kountz, Behnia,
  and Kapitulnik}}]{zhang_thermalization_2019}
\bibinfo{author}{\bibfnamefont{J.}~\bibnamefont{Zhang}},
  \bibinfo{author}{\bibfnamefont{E.~D.} \bibnamefont{Kountz}},
  \bibinfo{author}{\bibfnamefont{K.}~\bibnamefont{Behnia}}, \bibnamefont{and}
  \bibinfo{author}{\bibfnamefont{A.}~\bibnamefont{Kapitulnik}},
  \bibinfo{journal}{Proceedings of the National Academy of Sciences}
  \textbf{\bibinfo{volume}{116}}, \bibinfo{pages}{19869}
  (\bibinfo{year}{2019}), ISSN \bibinfo{issn}{0027-8424, 1091-6490},
  \urlprefix\url{https://www.pnas.org/content/116/40/19869}.

\bibitem[{\citenamefont{Ziman}()}]{Ziman_2001}
\bibinfo{author}{\bibfnamefont{J.~M.} \bibnamefont{Ziman}},
  \bibinfo{howpublished}{\underline{Electrons and Phonons: The Theory of
  }\newline \underline{Transport Phenomena in Solids} (Oxford University Press,
  1960)}.

\bibitem[{cla()}]{classical_scaling}
\bibinfo{howpublished}{The interatomic potential $V\left( \{r_i\} \right)$ can
  be rescaled into the form $V\left( \{r_i\} \right) = E_0 \overline{V}\left(
  \{r_i/a_0\} \right)$ where $\overline{V}$ is dimensionless, $E_0 = \hbar^2 /
  m a_0^2$, and $a_0$ is the Bohr radius. Assuming that the ion dynamics is
  classical, we obtain from dimensional analysis that $\tau_{\text{ph}} =
  \sqrt{\frac{M}{E_0}} a_0 f\left(T/E_0\right)$ with an unknown function $f$.
  Finally, assuming that at high temperatures $f \propto 1/T$, we find that
  $\tau_{\text{ph}} \sim \sqrt{\frac{M}{m}}\frac{\hbar}{T}$.}

\bibitem[{\citenamefont{Mousatov and Hartnoll}(2019)}]{mousatov_planckian_2019}
\bibinfo{author}{\bibfnamefont{C.~H.} \bibnamefont{Mousatov}} \bibnamefont{and}
  \bibinfo{author}{\bibfnamefont{S.~A.} \bibnamefont{Hartnoll}}
  (\bibinfo{year}{2019}), \bibinfo{note}{arXiv: 1908.04792},
  \urlprefix\url{http://arxiv.org/abs/1908.04792}.

\bibitem[{\citenamefont{Sachdev and Ye}(1993)}]{sachdev_gapless_1993}
\bibinfo{author}{\bibfnamefont{S.}~\bibnamefont{Sachdev}} \bibnamefont{and}
  \bibinfo{author}{\bibfnamefont{J.}~\bibnamefont{Ye}},
  \bibinfo{journal}{Physical Review Letters} \textbf{\bibinfo{volume}{70}},
  \bibinfo{pages}{3339} (\bibinfo{year}{1993}), ISSN \bibinfo{issn}{0031-9007},
  \bibinfo{note}{arXiv: cond-mat/9212030},
  \urlprefix\url{http://arxiv.org/abs/cond-mat/9212030}.

\bibitem[{\citenamefont{Kitaev}()}]{Kitaev_SYK_talk}
\bibinfo{author}{\bibfnamefont{A.}~\bibnamefont{Kitaev}},
  \bibinfo{howpublished}{A simple model of quantum holography.
  \url{http://online.kitp.ucsb.edu/online/entangled15/kitaev/} \newline
  \url{http://online.kitp.ucsb.edu/online/entangled15/kitaev2/}}.

\bibitem[{\citenamefont{Maldacena and Stanford}(2016)}]{maldacena_remarks_2016}
\bibinfo{author}{\bibfnamefont{J.}~\bibnamefont{Maldacena}} \bibnamefont{and}
  \bibinfo{author}{\bibfnamefont{D.}~\bibnamefont{Stanford}},
  \bibinfo{journal}{Physical Review D} \textbf{\bibinfo{volume}{94}}
  (\bibinfo{year}{2016}), ISSN \bibinfo{issn}{2470-0010, 2470-0029},
  \urlprefix\url{https://link.aps.org/doi/10.1103/PhysRevD.94.106002}.

\bibitem[{\citenamefont{Cugliandolo et~al.}(2001)\citenamefont{Cugliandolo,
  Grempel, and Santos}}]{cugliandolo_quantum_2001}
\bibinfo{author}{\bibfnamefont{L.~F.} \bibnamefont{Cugliandolo}},
  \bibinfo{author}{\bibfnamefont{D.~R.} \bibnamefont{Grempel}},
  \bibnamefont{and} \bibinfo{author}{\bibfnamefont{C.~A. d.~S.}
  \bibnamefont{Santos}}, \bibinfo{journal}{Physical Review B}
  \textbf{\bibinfo{volume}{64}}, \bibinfo{pages}{014403}
  (\bibinfo{year}{2001}), ISSN \bibinfo{issn}{0163-1829, 1095-3795},
  \bibinfo{note}{arXiv: cond-mat/0012222},
  \urlprefix\url{http://arxiv.org/abs/cond-mat/0012222}.

\bibitem[{\citenamefont{Edwards and Anderson}(1975)}]{edwards_theory_1975}
\bibinfo{author}{\bibfnamefont{S.~F.} \bibnamefont{Edwards}} \bibnamefont{and}
  \bibinfo{author}{\bibfnamefont{P.~W.} \bibnamefont{Anderson}},
  \bibinfo{journal}{Journal of Physics F: Metal Physics}
  \textbf{\bibinfo{volume}{5}}, \bibinfo{pages}{965} (\bibinfo{year}{1975}),
  ISSN \bibinfo{issn}{0305-4608},
  \urlprefix\url{https://doi.org/10.1088%2F0305-4608%2F5%2F5%2F017}.

\bibitem[{\citenamefont{Mezard et~al.}(1986)\citenamefont{Mezard, Parisi, and
  Virasoro}}]{mezard_spin_1986}
\bibinfo{author}{\bibfnamefont{M.}~\bibnamefont{Mezard}},
  \bibinfo{author}{\bibfnamefont{G.}~\bibnamefont{Parisi}}, \bibnamefont{and}
  \bibinfo{author}{\bibfnamefont{M.}~\bibnamefont{Virasoro}},
  \emph{\bibinfo{title}{Spin {Glass} {Theory} and {Beyond}}}, vol.
  \bibinfo{volume}{Volume 9} of \emph{\bibinfo{series}{World {Scientific}
  {Lecture} {Notes} in {Physics}}} (\bibinfo{publisher}{WORLD SCIENTIFIC},
  \bibinfo{year}{1986}).

\bibitem[{\citenamefont{Georges et~al.}(1999)\citenamefont{Georges, Parcollet,
  and Sachdev}}]{georges_mean-field_1999}
\bibinfo{author}{\bibfnamefont{A.}~\bibnamefont{Georges}},
  \bibinfo{author}{\bibfnamefont{O.}~\bibnamefont{Parcollet}},
  \bibnamefont{and} \bibinfo{author}{\bibfnamefont{S.}~\bibnamefont{Sachdev}},
  \bibinfo{journal}{arXiv:cond-mat/9909239}  (\bibinfo{year}{1999}),
  \bibinfo{note}{arXiv: cond-mat/9909239},
  \urlprefix\url{http://arxiv.org/abs/cond-mat/9909239}.

\bibitem[{\citenamefont{Baldwin and Swingle}(2019)}]{baldwin_quenched_2019}
\bibinfo{author}{\bibfnamefont{C.~L.} \bibnamefont{Baldwin}} \bibnamefont{and}
  \bibinfo{author}{\bibfnamefont{B.}~\bibnamefont{Swingle}}
  (\bibinfo{year}{2019}), \bibinfo{note}{arXiv: 1911.11865},
  \urlprefix\url{http://arxiv.org/abs/1911.11865}.

\bibitem[{\citenamefont{Facoetti et~al.}(2019)\citenamefont{Facoetti, Biroli,
  Kurchan, and Reichman}}]{facoetti_classical_2019}
\bibinfo{author}{\bibfnamefont{D.}~\bibnamefont{Facoetti}},
  \bibinfo{author}{\bibfnamefont{G.}~\bibnamefont{Biroli}},
  \bibinfo{author}{\bibfnamefont{J.}~\bibnamefont{Kurchan}}, \bibnamefont{and}
  \bibinfo{author}{\bibfnamefont{D.~R.} \bibnamefont{Reichman}},
  \bibinfo{journal}{Physical Review B} \textbf{\bibinfo{volume}{100}},
  \bibinfo{pages}{205108} (\bibinfo{year}{2019}),
  \urlprefix\url{https://link.aps.org/doi/10.1103/PhysRevB.100.205108}.

\bibitem[{\citenamefont{Chang et~al.}(2018)\citenamefont{Chang, Colin-Ellerin,
  and Rangamani}}]{chang_melonic_2018}
\bibinfo{author}{\bibfnamefont{C.-M.} \bibnamefont{Chang}},
  \bibinfo{author}{\bibfnamefont{S.}~\bibnamefont{Colin-Ellerin}},
  \bibnamefont{and} \bibinfo{author}{\bibfnamefont{M.}~\bibnamefont{Rangamani}}
  (\bibinfo{year}{2018}), \bibinfo{note}{arXiv: 1806.09903},
  \urlprefix\url{http://arxiv.org/abs/1806.09903}.

\bibitem[{\citenamefont{Giombi et~al.}(2017)\citenamefont{Giombi, Klebanov, and
  Tarnopolsky}}]{Giombi2017}
\bibinfo{author}{\bibfnamefont{S.}~\bibnamefont{Giombi}},
  \bibinfo{author}{\bibfnamefont{I.~R.} \bibnamefont{Klebanov}},
  \bibnamefont{and}
  \bibinfo{author}{\bibfnamefont{G.}~\bibnamefont{Tarnopolsky}},
  \bibinfo{journal}{Phys. Rev. D} \textbf{\bibinfo{volume}{96}},
  \bibinfo{pages}{106014} (\bibinfo{year}{2017}),
  \urlprefix\url{https://link.aps.org/doi/10.1103/PhysRevD.96.106014}.

\bibitem[{\citenamefont{Klebanov}()}]{Kleb_private}
\bibinfo{author}{\bibfnamefont{I.~R.} \bibnamefont{Klebanov}},
  \bibinfo{howpublished}{(private communication)}.

\bibitem[{\citenamefont{Azeyanagi et~al.}(2018)\citenamefont{Azeyanagi,
  Ferrari, and Massolo}}]{azeyanagi_phase_2018}
\bibinfo{author}{\bibfnamefont{T.}~\bibnamefont{Azeyanagi}},
  \bibinfo{author}{\bibfnamefont{F.}~\bibnamefont{Ferrari}}, \bibnamefont{and}
  \bibinfo{author}{\bibfnamefont{F.~I.~S.} \bibnamefont{Massolo}},
  \bibinfo{journal}{Physical Review Letters} \textbf{\bibinfo{volume}{120}},
  \bibinfo{pages}{061602} (\bibinfo{year}{2018}), ISSN
  \bibinfo{issn}{0031-9007, 1079-7114}, \bibinfo{note}{arXiv: 1707.03431},
  \urlprefix\url{http://arxiv.org/abs/1707.03431}.

\bibitem[{\citenamefont{Larkin and Ovchinnikov}(1969)}]{Larkin1969}
\bibinfo{author}{\bibfnamefont{A.}~\bibnamefont{Larkin}} \bibnamefont{and}
  \bibinfo{author}{\bibfnamefont{Y.~N.} \bibnamefont{Ovchinnikov}},
  \bibinfo{journal}{Sov Phys JETP} \textbf{\bibinfo{volume}{28}},
  \bibinfo{pages}{1200} (\bibinfo{year}{1969}).

\bibitem[{\citenamefont{Grozdanov et~al.}(2019)\citenamefont{Grozdanov, Schalm,
  and Scopelliti}}]{grozdanov_kinetic_2019}
\bibinfo{author}{\bibfnamefont{S.}~\bibnamefont{Grozdanov}},
  \bibinfo{author}{\bibfnamefont{K.}~\bibnamefont{Schalm}}, \bibnamefont{and}
  \bibinfo{author}{\bibfnamefont{V.}~\bibnamefont{Scopelliti}},
  \textbf{\bibinfo{volume}{99}}, \bibinfo{pages}{012206}
  (\bibinfo{year}{2019}), \bibinfo{note}{publisher: American Physical Society},
  \urlprefix\url{https://link.aps.org/doi/10.1103/PhysRevE.99.012206}.

\bibitem[{\citenamefont{Mao et~al.}(2019)\citenamefont{Mao, Chowdhury, and
  Senthil}}]{Mao_Chowdhury_Senthil_2019}
\bibinfo{author}{\bibfnamefont{D.}~\bibnamefont{Mao}},
  \bibinfo{author}{\bibfnamefont{D.}~\bibnamefont{Chowdhury}},
  \bibnamefont{and} \bibinfo{author}{\bibfnamefont{T.}~\bibnamefont{Senthil}}
  (\bibinfo{year}{2019}), \bibinfo{note}{arXiv: 1903.10499},
  \urlprefix\url{http://arxiv.org/abs/1903.10499}.

\bibitem[{\citenamefont{Cheng and Swingle}(2019)}]{cheng_chaos_2019}
\bibinfo{author}{\bibfnamefont{G.}~\bibnamefont{Cheng}} \bibnamefont{and}
  \bibinfo{author}{\bibfnamefont{B.}~\bibnamefont{Swingle}},
  \bibinfo{journal}{arXiv:1901.10446}  (\bibinfo{year}{2019}),
  \urlprefix\url{http://arxiv.org/abs/1901.10446}.

\bibitem[{\citenamefont{Song et~al.}(2017{\natexlab{a}})\citenamefont{Song,
  Jian, and Balents}}]{Balents}
\bibinfo{author}{\bibfnamefont{X.-Y.} \bibnamefont{Song}},
  \bibinfo{author}{\bibfnamefont{C.-M.} \bibnamefont{Jian}}, \bibnamefont{and}
  \bibinfo{author}{\bibfnamefont{L.}~\bibnamefont{Balents}},
  \bibinfo{journal}{Phys. Rev. Lett.} \textbf{\bibinfo{volume}{119}},
  \bibinfo{pages}{216601} (\bibinfo{year}{2017}{\natexlab{a}}),
  \urlprefix\url{https://link.aps.org/doi/10.1103/PhysRevLett.119.216601}.

\bibitem[{\citenamefont{Kitaev and Suh}(2018)}]{kitaev_soft_2018}
\bibinfo{author}{\bibfnamefont{A.}~\bibnamefont{Kitaev}} \bibnamefont{and}
  \bibinfo{author}{\bibfnamefont{S.~J.} \bibnamefont{Suh}},
  \bibinfo{journal}{Journal of High Energy Physics}
  \textbf{\bibinfo{volume}{2018}} (\bibinfo{year}{2018}), ISSN
  \bibinfo{issn}{1029-8479}, \bibinfo{note}{arXiv: 1711.08467},
  \urlprefix\url{http://arxiv.org/abs/1711.08467}.

\bibitem[{\citenamefont{Yeo and Moore}(2020)}]{yeo_does_2020}
\bibinfo{author}{\bibfnamefont{J.}~\bibnamefont{Yeo}} \bibnamefont{and}
  \bibinfo{author}{\bibfnamefont{M.~A.} \bibnamefont{Moore}},
  \bibinfo{journal}{arXiv:1911.02719 [cond-mat]}  (\bibinfo{year}{2020}),
  \bibinfo{note}{arXiv: 1911.02719},
  \urlprefix\url{http://arxiv.org/abs/1911.02719}.

\bibitem[{\citenamefont{Bray and Moore}(1980)}]{bray_replica_1980}
\bibinfo{author}{\bibfnamefont{A.~J.} \bibnamefont{Bray}} \bibnamefont{and}
  \bibinfo{author}{\bibfnamefont{M.~A.} \bibnamefont{Moore}},
  \bibinfo{journal}{Journal of Physics C: Solid State Physics}
  \textbf{\bibinfo{volume}{13}}, \bibinfo{pages}{L655} (\bibinfo{year}{1980}),
  ISSN \bibinfo{issn}{0022-3719},
  \urlprefix\url{https://doi.org/10.1088%2F0022-3719%2F13%2F24%2F005}.

\bibitem[{\citenamefont{Feynman and Hibbs}(1965)}]{feynman_quantum_1965}
\bibinfo{author}{\bibfnamefont{R.~P.} \bibnamefont{Feynman}} \bibnamefont{and}
  \bibinfo{author}{\bibfnamefont{A.~R.} \bibnamefont{Hibbs}},
  \emph{\bibinfo{title}{Quantum mechanics and path integrals}}
  (\bibinfo{publisher}{McGraw-Hill}, \bibinfo{year}{1965}).

\bibitem[{\citenamefont{Song et~al.}(2017{\natexlab{b}})\citenamefont{Song,
  Jian, and Balents}}]{song_strongly_2017}
\bibinfo{author}{\bibfnamefont{X.-Y.} \bibnamefont{Song}},
  \bibinfo{author}{\bibfnamefont{C.-M.} \bibnamefont{Jian}}, \bibnamefont{and}
  \bibinfo{author}{\bibfnamefont{L.}~\bibnamefont{Balents}},
  \bibinfo{journal}{Physical Review Letters} \textbf{\bibinfo{volume}{119}},
  \bibinfo{pages}{216601} (\bibinfo{year}{2017}{\natexlab{b}}), ISSN
  \bibinfo{issn}{0031-9007, 1079-7114}, \bibinfo{note}{arXiv: 1705.00117},
  \urlprefix\url{http://arxiv.org/abs/1705.00117}.

\bibitem[{\citenamefont{Kamenev}(2011)}]{kamenev_field_2011}
\bibinfo{author}{\bibfnamefont{A.}~\bibnamefont{Kamenev}},
  \emph{\bibinfo{title}{Field Theory of Non-Equilibrium Systems}}
  (\bibinfo{publisher}{Cambridge University Press}, \bibinfo{year}{2011}).

\end{thebibliography}

\appendix

\section{Replica analysis}
\label{sec:replica_formalism}

Here we give a few more details on the replica analysis. Starting from Eqn.~\ref{eq:replica_trick}, the replicated partition function is given by
\begin{equation}
    \overline{Z^n}=\int  \mathcal{D}{\boldsymbol{\phi}} \mathcal{D}\boldsymbol{v} \exp \left(-\sum_{\alpha=1}^{n} S_\alpha \right),
\label{eq:replicated_partition_function_of_phi}
\end{equation}
where $\boldsymbol{\phi}=\left\{ \phi_{i}^{\alpha}:i=1,...,N;\alpha=1,...,n\right\}  $, the disorder measure is 
\begin{equation}
    \mathcal{D}\boldsymbol{v}= \prod_{ijk}P\left(f_{ijk}\right) dv_{ijk} ,\text{  } P\left(v_{ijk}\right)=\frac{1}{\sqrt{4\pi v^{2}}}\exp\left(-\frac{v_{ijk}^{2}}{4v^{2}}\right),
\label{eq:disorder_measure}
\end{equation} and the action of each replica $\alpha=1,...,n$ is  
\begin{eqnarray}
S_\alpha &=& \int_{0}^{\beta}d\tau \Biggl( \sum_{i=1}^N \frac{1}{2}\phi_{i}^{\alpha} \left(-\partial_{\tau}^{2}+\Omega_{0}^{2} \right) \phi_{i}^{\alpha} 
         \\
      &+&  \frac{1}{N} \sum_{i,j,k}v_{ijk}\phi_{i}^{\alpha}\phi_{j}^{\alpha}\phi_{k}^{\alpha} + \frac{u}{4N} \left(\sum_{i=1}^N\left( \phi_{i}^{\alpha}\right)^{2} \right)^{2} \Biggr). \nonumber
\label{eq:action_of_phi}
\end{eqnarray}
To proceed, we integrate over the disorder and introduce composite fields $G_{\alpha\beta}\left(\tau,\tau'\right)$ and the and Lagrange multiplier fields enforcing these constraints, mentioned below Eqn.~\ref{eq:replica_trick}. The integration is straightforward and the implementation of the Lagrange multipliers is done with the identity \cite{kitaev_soft_2018}
\begin{eqnarray}
  f\left(\Xi\right)&=&\int_{-\infty}^{+\infty}dxf\left(x\right)\delta\left(x-\Xi\right) \nonumber\\
  &=&\frac{N}{2\pi}\int_{-\infty}^{+\infty}dx\int_{-\infty}^{+\infty}dyf\left(x\right)e^{iNy\left(x-\Xi\right)},
\end{eqnarray}

where $\Xi_{\alpha\beta}\left(\tau,\tau'\right)\equiv\frac{1}{N} \sum_a \phi_{a}^{\alpha}\left(\tau\right)\phi_{a}^{\beta}\left(\tau'\right)$ and $f\left(\Xi\right)=e^{\frac{Nv^{2}}{3}\Xi^{3}}$. This identity is enforced for all $\tau,\tau'$ that satisfy $(\tau,\tau')\in[0,\beta]^2 \text{ such that }\tau>\tau'$. This enforcement avoids the redundancy coming from the even parity of the imaginary-time Green's function. 

Then, we integrate out the phonon fields $\{\phi_i^{\alpha}\}$ and obtain that 
\begin{equation}
    \overline{Z^n} = \int \mathcal{D}\boldsymbol{G} \mathcal{D}\boldsymbol{\Pi} \exp\left(-nNS_{\text{eff}}\right),
\label{eq:replicated_partition_function_of_S_eff}
\end{equation}

where $\mathcal{D}\boldsymbol{A} = \prod_{\alpha\beta}\mathcal{D}A_{\alpha\beta}$ with $A=G,\Pi$, and the effective action is given by Eqn.~\ref{eq:S_eff_replicas}.

\subsection{Replica-diagonal solution}

The diagonal solution is given in and above Eqn.~\ref{eq:G_Pi_diagonal}. Here we discuss some other aspects related to the diagonal solution. 

\subsubsection{Instability of the disordered phase}

We comment on an instability that arises at intermediate temperatures, where the zero-frequency phonon stiffness $\hat{G}^{-1}(i\omega_n = 0)\equiv \Pi_0$ becomes negative. Given that the model is well-defined, in the sense that the energy is bounded from below, the existence of such instability is a first indication that the replica-diagonal solution is insufficient in this region of parameter space, suggesting the existence of a glass phase, as is later confirmed.

At high temperatures, the quartic term is dominant and the zero-frequency phonon stiffness is simply given by inverse of the positive renormalized phonon frequency. Problems may arise at low-to-intermediate temperatures, where the cubic term might become dominant. 

Let us focus on intermediate temperatures. We later show that this instability is can avoided at low temperatures as long as $\Omega_0$ is sufficiently large. To proceed, we solve the saddle-point Eqs.~\ref{eq:G_Pi_diagonal} for $\Pi_0$. We assume that $T$ is sufficiently large such that the Matsubara summation may be approximated by the $\omega_n=0$ component: 
\begin{eqnarray}
  \hat{\Pi}\left(0 \right) &=& v^{2}T\sum_{n\in\mathbb{Z}}\hat{G}\left(-i\omega_{n}\right)\hat{G}\left(i\omega_{n}\right)-uT\sum_{n\in\mathbb{Z}}\hat{G}\left(i\omega_{n}\right) \nonumber \\
  & \approx &  v^{2}T\hat{G}\left(0\right)^2-uT\hat{G}\left(0\right),
\end{eqnarray}
and substituting in $\Pi_0$ reads 
\begin{equation}
         \Pi_0^3 - \Omega_0^2\Pi_0^2 - u T \Pi_0 +v^2 T = 0.
    \label{eq:eq_for_Pi_0_rearng}
\end{equation}

The instability is characterized by parameters $\Omega_0,u,v$ and $T$ for which the only real solution of Eqn.~\ref{eq:eq_for_Pi_0_rearng} is negative.  
 We denote the ratio between the cubic and quartic energy scales as
\begin{equation}
    r \equiv \frac{\Omega_v}{\Omega_u} = \frac{v^{2/5}}{u^{1/3}}.
\end{equation}
To demonstrate the existence of an instability, we examine the cases where $r\to0$ and $r\gg1$. We first consider the $r\to 0$ case, which corresponds to setting $v\to0$ and $u>0$. Since the origin of this instability is the cubic term, this case is expected to show no instabilities. Indeed, we find that the real solution to Eqn.~\ref{eq:eq_for_Pi_0_rearng} for $T\gg u^{1/3}$ is given by
\begin{equation}
    \Pi_0 \approx \left(uT\right)^{1/2} > 0,
    \label{eq:pos_Pi_ex}
\end{equation}
where we have assumed for simplicity that $\Omega_0 \sim \Omega_u$. Note that this form of $\Pi_0$ also holds for $T\gg \Omega_u$ in the strongly coupled regime $\Omega_0 \sim \Omega_v \sim \Omega_u$.  

Now consider the case of  $r\gg1$, which corresponds to setting $u \to 0$ and $v>0$. Here we do expect an instability due to the fact that the cubic term becomes dominant at sufficiently high temperatures. Indeed, the only real solution of Eqn.~\ref{eq:eq_for_Pi_0_rearng} at $T\gg v^{2/5}$ is given by 
\begin{equation}
    \Pi_0 \approx -\left(v^2T\right)^{1/3},
    \label{eq:neg_Pi_ex}
\end{equation}
where we have assumed for simplicity that $\Omega_0 \sim v^{2/5}$. The requirement $\Pi_0>0$ is clearly violated, indicating that the replica-diagonal solution is unstable for $u\to0$. For large values of $r$, the instability exists at intermediate temperatures $v^{2/5} \lesssim T \lesssim rv^{2/5}$. In general, the instable region in parameter space for which $\Pi_0<0$ is also a function of $\Omega_0$, and can be characterized by solving the cubic polynomial in Eqn.~\ref{eq:eq_for_Pi_0_rearng} and demanding that the solution that is connected to the solution in Eqn.~\ref{eq:pos_Pi_ex} is positive for all $T\gtrsim\Omega_v$. However, the full characterization is not needed as the system undergoes a phase transition before it encounters this instability. This can be seen in Fig.~\ref{fig:single_band_pd}, where we find that smaller values of $u$ (larger values of $r$) correspond to a larger regions that realizes the glass phase.

At low temperatures, by approximating the Green's function as $\hat{G}(i\omega_n)^{-1} \approx \omega_n^2  + \Pi_0$, we find that 
\begin{eqnarray}
    \hat{\Pi}\left(0 \right) &=& v^{2}T\sum_{n\in\mathbb{Z}}\hat{G}\left(-i\omega_{n}\right)\hat{G}\left(i\omega_{n}\right)-uT\sum_{n\in\mathbb{Z}}\hat{G}\left(i\omega_{n}\right) \nonumber \\
  & \approx &  \frac{v^{2}}{4\Pi_0^{3/2}}-\frac{u}{\Pi_0^{1/2}},
\end{eqnarray}
 which gives the following equation for $\Pi_0$:
 \begin{equation}
     \Pi_0^{5/2} - \Omega_0^2 \Pi_0^{3/2} - u\Pi_0 + v^2/4=0.
     \label{eq:low_T_Pi_0}
 \end{equation}

One can check that the limit $u\to0$ still allows for positive solutions for $\Pi_0$ as long as $\Omega_0 \gtrsim v^{2/5}$, whereas for intermediate temperatures in the limit of $u\to 0$, the instability exists even for relatively large values of $\Omega_0$. Interestingly, for generic values of $\Omega_0,v$ and $u$, one may find multiple real and positive solutions for $\Pi_0$, from which one can understand the existence of multiple saddle points for the same set of parameters as mentioned briefly in \ref{relation_to_SYK}.

\label{sec:instability}

\subsubsection{Scale-invariant ansatz and relation to SYK}

\label{relation_to_SYK}

We show that the naive scaling ansatz is not a solution of the saddle-point equations. 
Recall that we have denoted the scaling ansatz by $G_\text{conf}\left( \tau \right) \equiv b/|\tau|^{2\Delta_{\phi}}$. By substituting the scaling ansatz in Eqs.~\ref{eq:G_Pi_diagonal_SYK}, one can check that $b^{-3}=3 v^{2} \left|\Gamma\left(-1/3\right)\right|\Gamma\left(1/3\right)$ and $\Delta_{\phi} = 1/3$. We assume that the scaling ansatz is valid up to a short-times cutoff $\Lambda^{-1}$. We then decompose the full two-point function as
\begin{equation}
    G\left(\tau\right) \equiv G_\text{conf}\left( \tau \right) \Theta\left( |\tau| - \Lambda^{-1} \right) +  G_{\text{UV}}\left( \tau \right) \Theta\left( \Lambda^{-1} - |\tau| \right)
\end{equation}
where $\Theta(x) = 1$ if $x>0$ and zero otherwise, and the short-times piece is given by $G_{\text{UV}}\left( \tau \right) =\int_{\left|\omega\right|>\Lambda}\frac{d\omega}{2\pi}\frac{e^{-i\omega\tau}}{\omega^{2}} $. In the fermionic case, the contribution of this short-times piece to the self-energy may be neglected due the odd parity of the fermionic Green's function. Here this is not the case. Instead, it contribute as a constant, which implies that our naive approach, where we took $\Omega_0 \to 0$ to obtain a set of `SYK-like' saddle-point equations (Eqs.~\ref{eq:G_Pi_diagonal_SYK}), is inconsistent. 

The short-times piece may approximated as $G_{\text{UV}}\left( \tau \right) \approx 1/\pi \Lambda$, since we are interested in the long-times behavior of $G$. Then, the self-energy reads 
\begin{eqnarray}
 \Pi\left(\tau\right) &=& v^2 G\left(\tau\right)^{2} \nonumber \\
                      &=& v^{2}G_\text{conf}\left(\tau\right)^{2} \Theta\left( |\tau| - \Lambda^{-1} \right) \nonumber \\
                      &+& v^{2}\left(\frac{1}{\pi\Lambda}\right)^{2} \Theta\left( \Lambda^{-1} - |\tau| \right).
\end{eqnarray}
Then, up to corrections of order $\omega/\Lambda$, we have that 
\begin{equation}
     \Pi\left(i\omega\right) = \Pi_\text{conf} \left(i\omega\right) +  \underbrace{\frac{2v^{2}}{\pi^{2}\Lambda^{3}}}_{\equiv \Pi_{\text{UV}}},
\end{equation}
where $\Pi_\text{conf} \left(i\omega\right) = - G_\text{conf} \left(i\omega\right)^{-1} $. 

Reinstating $\Omega_0\to \Omega_\text{conf}\equiv\Pi_{\text{UV}}$, we see that
\begin{eqnarray}
     G^{-1}(i\omega) &=& \omega^2 + \Omega_\text{conf}^2 - \Pi(i\omega) \nonumber \\
                    &\approx&\omega^2 + \Pi_{\text{UV}} -\Pi_\text{conf} \left(i\omega\right) -  \Pi_{\text{UV}} \nonumber \\
                    &\approx & - \Pi_\text{conf} \left(i\omega\right)
\end{eqnarray} 
in the long-times limit. The cutoff can be found by the consistency requirement: $G_{\text{UV}}^{-1}(i\omega = i\Lambda) \approx G_\text{conf}^{-1}(i\omega = i\Lambda)$, which then enables us to extract $\Omega_\text{conf} \approx v^{2/5}$. $\Omega_\text{conf}$ can also be extracted numerically from $\delta\Pi = \Omega_0^2 + \Pi_\text{conf}(i\omega_n=0)$ (see below Eqn.~\ref{eq:modified_updating_scheme}), we find good agreement as $\Omega_\text{conf} \approx 1.1 v^{2/5}$ for $v^{2/5}\beta \gg 1$. 

For completeness let us note that the mapping from $T=0$ to $T>0$ is identical to the fermionic SYK case (see e.g. \cite{maldacena_remarks_2016}) with the scaling dimension $\Delta_\phi = 1/3$. At finite $T$, 
\begin{equation}
\label{conformal_G}
    \hat{G}(i\omega_n) = \tilde{b}\frac{\beta^{\Delta_\phi}}{v^{2\Delta_\phi}}\frac{\Gamma\left(\Delta_\phi + \frac{\beta\omega_n}{2\pi}\right)}{\Gamma\left(1 - \Delta_\phi + \frac{\beta\omega_n}{2\pi}\right)}
\end{equation}
where $\tilde{b} = (2\pi)^{2\Delta_\phi}/\Gamma(2\Delta_\phi)\left(3 \left|\Gamma\left(-1/3\right)\right|\Gamma\left(1/3\right)\right)^{\Delta_\phi}$.

We have shown that the model may be fine-tuned to a conformally-invariant critical point by a deformation of the bare phonon frequency $\Omega_0\to\Omega_\text{conf}$. As noted in Sec.~\ref{relation_to_syk_main_text}, however, the disordered saddle-point equations admits a different, gapped solution for this set of parameters ($u\to0$, $\Omega_0\to\Omega_\text{conf}$) which is thermodynamically favorable.  

\subsection{1-step replica symmetry breaking solution}

As shown in \cite{cugliandolo_quantum_2001}, the glass phase of the model is described by a 1-step replica symmetry breaking (1SRSB) solution at the level of the saddle-point approximation\footnote{See \cite{yeo_does_2020} for a recent study of the replica structure of $p$-spin models beyond the saddle-point approximation.}. The following derivation is largely along the lines of \cite{cugliandolo_quantum_2001}. 
We begin from the 1SRSB solution, 
\begin{equation}
    {G}_{\alpha\beta}\left( \tau \right) = \left( {g}_{d}\left( \tau \right)-{g}_{EA} \right)\delta_{\alpha\beta}+\left({g}_{EA} - {g}_{0}\right)\epsilon_{\alpha\beta} + {g}_{0},    
\label{eq:1SRSB_solution}
\end{equation}
where $\epsilon_{\alpha\beta}=1$ if $\alpha$ and $\beta$ are in a diagonal block of size $m$ and $\epsilon_{\alpha\beta}=0$ otherwise. Interestingly, the off-diagonal terms of ${G}_{\alpha\beta}\left( \tau \right)$ can be shown to be $\tau$-independent \cite{bray_replica_1980}.

As in \cite{cugliandolo_quantum_2001}, the absence of a linear-in-$\phi_{i}$ term in $H$ implies that $g_{0} = 0$. To proceed, we preform the following steps. We substitute the the self-energy, obtained by the variation of $S_{\text{eff}}$ with respect to $\Pi_{\alpha\beta}\left(\tau\right)$, back in $S_{\text{eff}}$. Then we move to Matsubara space. And lastly we substitute the (replica space) eigenvalues and corresponding degeneracies of $\hat{G}_{\alpha\beta}\left(i\omega_n\right)$, which are given in terms of  $\hat{g}_d\left(i\omega_n\right),\hat{g}_{EA}$ and $m$, in the $\ln\text{det}$ term. These steps will allow us to easily take the limit $n\to0$, where $m$ is then analytically continued to take real values between zero and one such that $m=1$ corresponds to the replica diagonal solution and $m=0$ to the replica symmetric solution. We continue by scaling dimensionful quantities with respect to $\Omega_v = v^{2/5}$, and to lighten the notation, we leave the dimensionless parameters with the same notation. So in practice, we simply set $v=1$ wherever it appears and remember that all other parameters are scaled with respect to the appropriate power of $v$. The saddle-point equations are then given by 
\begin{eqnarray}
0 &=& \frac{\hat{g}_{d}\left(i\omega_{n}\right)+\left(m-2\right)\hat{g}_{EA}}{\left(\hat{g}_{d}\left(i\omega_{n}\right)-\hat{g}_{EA}\right)\left(\hat{g}_{d}\left(i\omega_{n}\right)+\left(m-1\right)\hat{g}_{EA}\right)} \nonumber \\ 
&-& \omega_{n}^{2} - \Omega_{0}^2 + \hat{\Pi}\left(i\omega_{n}\right),
\label{eq:SD_gd} \\
0 &=& \frac{1}{\left(\hat{g}_{d}\left(0\right)-\beta g_{EA}\right)\left(\hat{g}_{d}\left(0\right) + \left(m-1\right)\beta g_{EA}\right)} -g_{EA}, \nonumber \\
 & & 
 \label{eq:SD_gEA} \\
0 &=& \frac{1}{m}\frac{\beta g_{EA}}{\hat{g}_{d}\left(0\right)+\left(m-1\right)\beta g_{EA}}+ \frac{1}{3}\beta^{2}g_{EA}^{3}  \nonumber \\ 
  &+& \frac{1}{m^{2}}\ln\left(\frac{\hat{g}_{d}\left(0\right)-\beta g_{EA}}{\hat{g}_{d}\left(0\right)+\left(m-1\right)\beta g_{EA}}\right),
\label{eq:SD_m}
\end{eqnarray}

where the self-energy is given by ${\Pi}\left( \tau \right) =  {g}_{d}(\tau)^2 -ug_{d}\left( \tau \right)\delta(\tau)$.
Here Eqs.~\ref{eq:SD_gd},\ref{eq:SD_gEA} and \ref{eq:SD_m} are obtained by varying $S_{\text{eff}}$ with respect to $\hat{g}_d\left(i\omega_{n}\right)$, $\hat{g}_{EA}$ and $m$, respectively. Note that we have already eliminated solutions for which $g_{EA}=0$ or $m=1$ in the saddle-point equations.

Following \cite{cugliandolo_quantum_2001}, we define two parameters, $y \equiv \beta g_{EA} / \hat{g}_{d}\left(0\right)$ and $x \equiv my / \left( 1-y \right) $. Substituting these in Eqn.~\ref{eq:SD_gEA} and Eqn.~\ref{eq:SD_m} gives an equation for $x$, 
\begin{equation}
    0=\ln\left(\frac{1}{1+x}\right)+\frac{x}{1+x}+\frac{1}{3}\frac{x^{2}}{1+x}.
\label{eq:eq_for_x}    
\end{equation}
Numerically solving Eqn.~\ref{eq:eq_for_x} gives $x=1.81696$. It will be useful to notice that
\begin{equation}
    g_{EA}^{3}=\frac{1}{m^{2}\beta^{2}}\frac{x^{2}}{1+x} \quad,\quad \hat{g}_{d}\left(0\right) = \beta g_{EA}\frac{x+m}{x}.
\label{eq:relation_for_gd_and_gEA}
\end{equation} In particular, note that $\hat{g}_{d}\left(0\right)$ is fixed by $m$ and $\beta$. We then separate $g_d$ and the self-energy to a constant and $\tau$-dependent parts, 
\begin{eqnarray}
g_d\left( \tau \right) &\equiv& g_{EA} + \tilde{G}\left( \tau \right), 
\label{eq:gd_to_G_tilde} \\
\Pi \left( \tau \right) &\equiv& g_{EA}^2 + \tilde{\Pi} \left( \tau \right).
\label{eq:Pi_to_Pi_tilde}
\end{eqnarray}
Substituting Eqn.~\ref{eq:gd_to_G_tilde} and Eqn.~\ref{eq:Pi_to_Pi_tilde} into Eqn.~\ref{eq:SD_gd}, with the help of the relations in Eqn.~\ref{eq:relation_for_gd_and_gEA}, the terms proportional to $\delta_{\omega_n,0}$ cancel and we get $\hat{\tilde{G}}\left(i\omega_{n}\right)^{-1}= \omega_{n}^{2}+\Omega_{0}^2-\hat{\tilde{\Pi}}\left(i\omega_{n}\right)$.

It is important to remember to respect the saddle-point constraint on  $\hat{g_d}\left(0\right)$, which implies $\hat{\tilde{G}}\left(0\right) = m\beta g_{EA}/x$. Therefore, for a given $m$ and $\beta$, $\Omega_{0}$ must be chosen self-consistently such that the constraint is satisfied. That is, we are forced to set
\begin{equation}
    \Omega_{0}^2 = \frac{x}{m\beta g_{EA}} + \hat{\tilde{\Pi}}\left(0\right).
\label{eq:self_consistent_omega_0}
\end{equation}

Finally, we arrive at a closed set of equations for $\tilde{G}$ and $\tilde{\Pi}$,
\begin{eqnarray}
    \hat{\tilde{G}}\left(i\omega_{n}\right) &=& \frac{1}{ \omega_{n}^{2} +   \frac{x}{m\beta g_{EA}}-\hat{\tilde{\Pi}}\left(i\omega_{n}\right)+\hat{\tilde{\Pi}}\left(0\right)}
\nonumber, \\
    {\tilde{\Pi}}\left(\tau\right) &=& {\tilde{G}}\left(\tau \right)^2+2g_{EA}{\tilde{G}}\left(\tau \right) \nonumber \\
    &-& u\left(\tilde{G}\left(\tau\right)+g_{EA}\right)\delta(\tau).
\label{eq:Pi_tilde_1SRSB} 
\end{eqnarray}
These equations can be solved numerically (see Appendix~\ref{sec:numerical_methods}) to obtain the free-energy density of the glass phase and construct the phase diagram of the model. In practice we fix $m$ and $\beta$, numerically solve for $\hat{G},\hat{\Pi}$ and then extract $\Omega_{0}$ from Eqn.~\ref{eq:self_consistent_omega_0}. 

\subsection{Thermodynamic functions}
\label{td_functions}
The thermodynamic functions (free energy, internal energy and specific heat) of the model are obtained from the effective action in Eqn.~\ref{eq:S_eff_replicas}. The free energy of model is obtained by substituting the definitions of the replica space solutions into Eqn.~\ref{eq:S_eff_replicas}. Then, we find that the free energy of the 1SRSB solution is given by
\begin{widetext}
\begin{eqnarray}
  2\beta \overline{f}	&=&-\left(\frac{m-1}{m}\right)\ln\left(\frac{1-y}{1-\left(1-m\right)y}\right)-\ln\left(1+\left(m-1\right)y\right)-\sum_{n}\ln\left(\left(\omega_{n}^{2}+\Omega_{0}^{2}\right)\hat{g}_{d}\left(i\omega_{n}\right)\right)
	\nonumber\\
	&+&\sum_{n}\left(\left(\omega_{n}^{2}+\Omega_{0}^{2}\right)\hat{g}_{d}\left(i\omega_{n}\right)-1\right)-\frac{v^2}{3}\left(\beta\int_{0}^{\beta}d\tau g_{d}^{3}\left(\tau\right)+\left(m-1\right)\beta^{2}g_{EA}^{3}\right)+\frac{u}{2}\beta g_{d}\left(\tau=0\right)^{2}+C,	
	\label{eq:free_energy_glass}
\end{eqnarray}
\end{widetext}
and for the diagonal solution, which corresponds to taking the limit $m\to1$ or the limit $g_{EA}\to0$, we have 
\begin{eqnarray}
  \label{eq:free_energy_dis}
  2\beta\overline{f} &=&-\sum_{n}\ln\left(\left(\omega_{n}^{2}+\Omega_{0}^{2}\right)\hat{G}\left(i\omega_{n}\right)\right)\\
  &+&\sum_{n}\left(\left(\omega_{n}^{2}+\Omega_{0}^{2}\right)\hat{G}\left(i\omega_{n}\right)-1\right)\nonumber\\
  &-&\frac{\beta v^2}{3}\int_{0}^{\beta}d\tau G^{3}\left(\tau\right)+\frac{\beta u}{2} G\left(\tau=0\right)^{2}+C. \nonumber 
\end{eqnarray}
Here, $C$ is related to the regularization of the $\ln\det$ term, $C=\sum_n \ln\left(\beta^2 \left(\omega_n^2 + \Omega_0^2\right)\right)$. It can be shown that $C=2\ln\left(2\sinh\left(\frac{\beta\Omega_0}{2}\right)\right)$  \cite{feynman_quantum_1965} (which is simply the free energy of an harmonic oschillator with frequency $\Omega_0$). To derive the internal energy of the disordered phase, however, one can use the unregularized form of the free energy, as we show next.

The internal energy $U$ is defined by $\partial (\beta f) / \partial \beta$. Due to the saddle-point approximation, the differentiation is non-vanishing only when it acts explicitly on $\beta$:
\begin{eqnarray*}
    U &=& \frac{\partial \left( \beta f \right)}{\partial \beta} 
    =\frac{\partial  S_\text{eff}(\beta,G,\Pi) }{\partial \beta} \nonumber \\
    &=&  \frac{\partial  S_\text{eff} }{\partial \beta}  + \frac{\delta  S_\text{eff} }{\delta G}\frac{\partial  G }{\partial \beta}+ \frac{\delta  S_\text{eff} }{\delta \Pi}\frac{\partial  \Pi }{\partial \beta} \nonumber \\ 
    &=&  \frac{\partial  S_\text{eff} }{\partial \beta} .
    \label{stationarity_cond}
\end{eqnarray*}

To proceed, we substitute the saddle-point Eqs.~\ref{eq:G_Pi_diagonal} in Eqn.~\ref{eq:free_energy_dis} and obtain that
\begin{eqnarray*}
    \beta\overline{f} &=& {\frac{1}{2}\sum_{n}\ln\left(\beta^2 \left(\omega_{n}^2 + \Omega_0^2 - \hat{\Pi}(i\omega_n) \right)\right)}
  \nonumber \\
  &+& \frac{\beta v^2}{3}\int_{0}^{\beta}d\tau G^{3}\left(\tau\right)-\frac{\beta u}{4} G\left(\tau=0\right)^{2}.
  \label{eq:free_energy_dis_not_simp}
\end{eqnarray*}
The derivative of the first term is evaluated as
\begin{eqnarray*}
     &&\partial_\beta \left( \frac{1}{2} \sum_n  \ln \left( \overline{\omega}_n^2 + \beta^2\left( \Omega_0^2 - \hat{\Pi}(i\omega_n) \right) \right) \right)\nonumber \\
    &=&  \frac{1}{2} \sum_n  \frac{2 \beta\left( \Omega_0^2 - \hat{\Pi}(i\omega_n) \right)}{\left( \overline{\omega}_n^2 + \beta^2\left( \Omega_0^2 - \hat{\Pi}(i\omega_n) \right) \right) }\nonumber \nonumber \\
     &=& \frac{1}{\beta} \sum_n  \left( \Omega_0^2\hat{G}(i\omega_n) - \hat{\Pi}(i\omega_n)\hat{G}(i\omega_n) \right)\nonumber \\
     &=& \Omega_0^2{G}(\tau=0) - \int_0^\beta  d \tau {G}(\tau)\Pi(\tau),\nonumber \\
\end{eqnarray*}
where $\overline{\omega_n} \equiv 2\pi n$.
To differentiate the first summand in the second line, we may rescale the intergration variable $\tau\to x \equiv \tau / \beta$, similarly to \cite{cugliandolo_quantum_2001}. Finally, we obtain that 
\begin{equation}
     U = \Omega_0^2{G}(\tau=0) - \frac{v^2}{3}\int_{0}^{\beta}d\tau G^{3}\left(\tau  \right) + \frac{3u}{4} G\left(\tau=0\right)^{2}.
     \label{internal_E}
\end{equation}

Let us use Eqn.~\ref{internal_E} to evaluate the specific heat $c \equiv  \partial U / \partial T$. We will show the high-temperature specific heat. At low temperatures the evaluation is similar and one finds that the $c$ vanishes exponentially as $T\to0$ since the system is gapped at $T=0$. 

For $T\gg \Omega_u$, we use Eqn.~\ref{eq:pos_Pi_ex} to approximate the Green's function as $G(\tau) \approx \sqrt{T/u}$. Substituting this approximation in Eqn.~\ref{internal_E}, we obtain the high-temperature specific heat, given in Eqn.~\ref{c_high_T}. The sign of prefactor $b$ determines whether $c$ has a maximum at an intermediate temperature ($b>0$) or saturates to its maximum at $T/\Omega_u \to \infty$ ($b<0)$. 

Moreover, note that the high-temperature limit 
\begin{equation}
    \lim_{T/\Omega_u \to \infty} c = \frac{3}{4} 
\end{equation}
is not an artifact of the saddle-point approximation. It can be derived by a simple scaling argument: Consider a classical system (with position $X$ and momentum $P$) where the highest order term in the potential is quartic, and denote it by $u X^4$. At sufficielty high temperature, the quartic term is dominant. The partition function of the system may thus be approximation as
\begin{equation}
    Z \approx \left( \int dP e^{-\beta P^2/2M} \right)\left( \int dX e^{-\beta u X^4} \right).
\end{equation}
By rescaling 
\begin{eqnarray*}
X & \to & \overline{X} = X/(\beta u)^{1/4} ;\nonumber \\
P & \to & \overline{P} = P/(\beta/2M)^{1/2}, 
\end{eqnarray*}
we find that  
\begin{eqnarray*}
    Z &\approx& \left((\beta/2M)^{-1/2} \int d\overline{P} e^{- \overline{P}^2} \right)\left( (\beta u)^{-1/4} \int d\overline{X} e^{-\overline{X}^4} \right) \nonumber \\
    &=& T^{3/4}\times(\text{terms independent of $T$}).
\end{eqnarray*}
Hence the free energy is given by 
\begin{equation}
    \beta F = \ln(Z) = \frac{3}{4}\ln(T) + \ln(\text{terms independent of $T$})
\end{equation}
and it follows that $c \to 3/4$ as $T\to\infty$.

\section{Keldysh formalism}
\label{sec:keldysh_formalism}

We derive the Keldysh action and saddle-point equations of the disordered phase to study the dynamical properties of the phonons, and in particular the phonon lifetime. The derivation below is done in the spirit of \cite{song_strongly_2017}. In this formalism we calculate the partition function $Z=\text{Tr}\left(e^{-\beta H}U\right)/\text{Tr}\left(e^{-\beta H}\right)$ where $U$ is the identity real-time evolution operator, evolving forward from time $t_0$ to $t_f$ and backward from $t_f$ to $t_0$. As usual, the label $+$ ($-$) denotes the Keldysh forward (backward) contour. 

The disorder-averaged Keldysh partition function $\overline{Z}$ is given by  
\begin{equation}
    \overline{Z} = \int \mathcal{D}\boldsymbol{\phi}  \mathcal{D}\boldsymbol{v} \exp\left( i S_K \right),
\end{equation}
where  $\boldsymbol{\phi}=\left\{ \phi_{i}:i=1,...,N\right\}$, the disorder measure was defined in Eqn.~\ref{eq:disorder_measure}, and the Keldysh action is 
\begin{eqnarray}
S_K &=& \sum_{s=\pm1} s \int_{t_{0}}^{t_{f}}dt \Biggl( \frac{1}{2}\sum_{i=1}^N\phi_{i,s}\left(-\partial_{t}^{2} - \Omega_{0}^{2} \right)\phi_{i,s}  \\
    &-& \frac{1}{N}\sum_{i,j,k}v_{ijk}\phi_{i,s}\phi_{j,s}\phi_{k,s} -  \frac{u}{4N} \left(\sum_{i=1}^N \phi_{i,s}^{2} \right)^{2} \Biggr).    \nonumber
\label{eq:Keldysh_action_of_phi}    
\end{eqnarray}

Similarly to the replica method, we average over the disorder and introduce composite fields
\begin{equation}
    G_{ss'}\left(t,t'\right)=-\frac{i}{N}\sum_{i=1}^{N}\phi_{i,s}\left(t\right)\phi_{i,s'}\left(t'\right)
\label{eq:Keldysh_contour_Green_function_def}
\end{equation}
and Lagrange multiplier fields enforcing these constraints $\Pi_{ss'}\left(t,t'\right)$. We then integrate out the phonon fields to obtain the disorder-averaged partition function in terms of $G$ and $\Pi$,
\begin{equation}
    \overline{Z} = \int \mathcal{D}G\mathcal{D}\Pi\exp\left(iN\bar{S}_{K}\right),
\end{equation}
where the effective Keldysh action is given by 
 \begin{widetext}
\begin{eqnarray}
 i\bar{S}_K  &=& -\frac{1}{2}\ln\det\left(s\delta_{ss'}\delta\left(t-t'\right)\left(-\partial_{t}^{2}-\Omega_{0}^2 \right)-\Pi_              {ss'}\left(t,t'\right)\right) \\
    &+& \frac{1}{2}\sum_{s,s'=\pm1}\int_{t_{0}}^{t_{f}}dtdt'\Biggl(iss'\frac{v^{2}}{3}G_{ss'}\left(t,t'\right)^{3} 
              - is\frac{u}{2}G_{ss'}\left(t,t'\right)^{2} \delta\left(t-t'\right)\delta_{ss'} \nonumber - \Pi_{ss'}\left(t,t'\right)G_{ss'}\left(t,t'\right) \Biggr) . \nonumber        
\label{eq:S_eff_Keldysh}
\end{eqnarray}
              
\end{widetext}

The saddle-point equation are then obtained by varying the effective Keldysh action with respect to $G_{ss'}$ and $\Pi_{ss'}$,
\begin{eqnarray}
\label{eq:saddle_point_before_Keldysh_rot}
\hat{G}_{ss'}\left(\omega\right) &=& \left(s\delta_{ss'}\left(\omega^2-\Omega_{0}^2 \right)-\hat{\Pi}_{ss'}\left(\omega \right)\right)^{-1} , \\
\Pi_{ss'}\left(t\right) &=& iss' v^{2} G_{ss'}\left(t\right)^{2} 
              - i s u G_{ss'}\left(t\right) \delta\left(t\right)\delta_{ss'} \nonumber,
\end{eqnarray}
where we have assumed real-time translation invariance.

To obtain the saddle-point equations in terms of the more conventional retarded, advanced and Keldysh Green's functions, we introduce a Keldysh rotation matrix, 
\begin{equation}
    L = \frac{1}{\sqrt{2}}\begin{pmatrix} 
1 & 1 \\
1 & -1 
\end{pmatrix},
\label{eq:rot_mat_Keldysh}
\end{equation}
which satisfies
\begin{equation}
     \begin{pmatrix} 
G_K & G_R \\
G_A & 0 
\end{pmatrix} = L^{\dagger} \begin{pmatrix} 
G_{++} & G_{+-} \\
G_{-+} & G_{--} 
\end{pmatrix} L,
\label{eq:Keldysh_rot_G}
\end{equation} and 
\begin{equation}
     \begin{pmatrix} 
0 & \Pi_A \\
\Pi_R & \Pi_K 
\end{pmatrix} = L^{\dagger} \begin{pmatrix} 
\Pi_{++} & \Pi_{+-} \\
\Pi_{-+} & \Pi_{--} 
\end{pmatrix} L.
\label{eq:Keldysh_rot_Pi}
\end{equation}

The saddle-point Eqs.~\ref{eq:saddle_point_before_Keldysh_rot} do not contain the information about the initial thermal equilibrium density matrix. Thermal equilibrium at inverse temperature $\beta$ is imposed by setting the Keldysh function according to the flactuation-dissipation theorem, $\hat{G}_K(\omega) =\coth\left(\beta\omega/2\right)\left(\hat{G}_{R}\left(\omega\right)-\hat{G}_{A}\left(\omega\right)\right)$. Using equations Eqn.~\ref{eq:saddle_point_before_Keldysh_rot}, Eqn.~\ref{eq:Keldysh_rot_G} and Eqn.~\ref{eq:Keldysh_rot_Pi} and the fact that $\hat{G}_{R}^*\left(\omega\right)=\hat{G}_{A}\left(\omega\right)$, we obtain the Keldysh saddle-point Eqs.~\ref{eq:Keldysh_sc_eqs}.

The generalization to the MB model is given by

\begin{eqnarray}
 \hat{G}_{aR}\left(\omega\right) &=& -\frac{1}{\omega^{2}-\Omega_{a}^{2}-\hat{\Pi}_{R}\left(\omega\right)}  , \\
 \hat{G}_{aK}\left(\omega\right)&=&2i\coth\left(\frac{\beta\omega}{2}\right)\text{Im}\left[\hat{G}_{aR}\left(\omega\right)\right], \nonumber \\
 \Pi_R\left(t\right) &=& i\frac{v^{2}}{N^2}\sum_{a,b}{G}_{aR}\left( t \right) {G}_{bK}\left( t \right) -\frac{iu}{2N}\sum_a G_{aK}\left(t\right)\delta(t) \nonumber, 
 \label{eq:Keldysh_many_bands_sc_eqs}
\end{eqnarray}
where $a,b=1,...,N$. Note that this approach is equivalent to the Keldysh diagrammatic approach when considering the $\mathcal{O}(1)$ diagrams in a $1/N$ expansion, see e.g. \cite{kamenev_field_2011}. 

\subsection{Real-time derivation of $\Pi_0$ at high $T$}

\label{sec:real_time_Pi_0}

In Section~\ref{sec:instability}, we derived an equation for $\Pi_0$ in imaginary-time, where the high-temperature limit allowed us to approximate the Matsubara summation by the zeroth Matsubara frequency. Here we show that Eqn.~\ref{eq:eq_for_Pi_0_rearng} can be derived exactly by solving the Keldysh equations for $\hat{\Pi}_R(0)$. 

To begin, we use Eqs.~\ref{eq:Keldysh_sc_eqs} and substitute the Keldysh Green's function in $\hat{\Pi}_R(0)$, 

\begin{eqnarray}
   \hat{\Pi}_R\left(0\right) &=& - 2v^{2} \int \frac{d\omega}{2\pi} \hat{G}_{R}(\omega) \coth\left(\frac{\beta\omega}{2}\right)\text{Im}\left[\hat{G}_{R}\left(\omega\right)\right] \nonumber \\ &+& u \int \frac{d\omega}{2\pi} \coth\left(\frac{\beta\omega}{2}\right)\text{Im}\left[\hat{G}_{R}\left(\omega\right)\right]
   \label{eq:real_time_Pi_0_before_classical_FDT}
\end{eqnarray}
where we have used the even parity of $\hat{G}_K$. 

To proceed, we expand the $\coth$ to leading order, thereby replacing the quantum with its classical version. We later show that this is consistent for $T \gg \Omega_u$. We have 
\begin{eqnarray}
   \hat{\Pi}_R\left(0\right) &=& - 4v^{2}T \int \frac{d\omega}{2\pi} \hat{G}_{R}(\omega) \frac{\text{Im}\left[\hat{G}_{R}\left(\omega\right)\right]}{\omega} \nonumber \\ &+&  u T \int \frac{d\omega}{\pi} \frac{\text{Im}\left[\hat{G}_{R}\left(\omega\right)\right]}{\omega}.
   \label{eq:real_time_Pi_0}
\end{eqnarray}
Using the Kramers-Kronig relations, the second summand in Eqn.~\ref{eq:real_time_Pi_0} is given by
\begin{equation}
     uT \text{Re}\left[\hat{G}_R(0)\right] = uT \hat{G}_R(0).
\end{equation} 

Denote 
\begin{equation}
    I \equiv \int \frac{d\omega}{2\pi} \hat{G}_{R}(\omega) \frac{\text{Im}\left[\hat{G}_{R}\left(\omega\right)\right]}{\omega}.
\end{equation}
In the following we show that 
\begin{equation}
    I= \frac{1}{4}\hat{G}_R(0)^2.
    \label{eq:identity_goal}
\end{equation}
Then, by substituting the two summands back to Eqn.~\ref{eq:real_time_Pi_0}, together with the fact that $\Pi_0 = \hat{G}_R(0)^{-1}$, we obtain Eqn.~\ref{eq:eq_for_Pi_0_rearng} from the Keldysh equations.

To derive Eqn.~\ref{eq:identity_goal}, we rewrite $I$ as
\begin{equation}
    I = \int \frac{d\omega}{2\pi} \hat{G}_R(\omega) \frac{\hat{G}_R(\omega) - \hat{G}_A(\omega)}{2i\omega}.
\end{equation}
Using the fact that 
\begin{equation}
    \int \frac{d\omega}{2\pi} \frac{e^{-i\omega t}}{i\omega} = -\frac{1}{2}\text{sgn}(t),
\end{equation}
we can write $I$ as 
\begin{eqnarray}
  I &=& - \int \frac{d\omega}{2\pi} \int dt_1dt_2dt_3 e^{-i\omega(t_1 + t_2 + t_3)} \nonumber \\ 
    & & \times \frac{1}{4} G_R(t_1) \left[G_R(t_2) - G_A (t_2) \right] \text{sgn}(t_3)  \nonumber \\
    &=& \int dt_1 dt_2 \frac{1}{4} G_R(t_1) \left[G_R(t_2) - G_A (t_2) \right] \nonumber \\
    & & \times \text{sgn}(t_1 + t_2).
\end{eqnarray}
Expanding the square brackets, the first term is 
\begin{eqnarray}
  & & \int dt_1 dt_2 \frac{1}{4}  G_R(t_1) G_R(t_2)  \text{sgn}(t_1 + t_2) \nonumber \\
  &=& \int dt_1 dt_2 \frac{1}{4}  G_R(t_1) G_R(t_2)  \nonumber \\
    &=&  \frac{1}{4}  \hat{G}_R(0)^2 .
\end{eqnarray}
This is because $G_R(t) = 0$ for $t<0$, and hence the sign functin can be replaced by unity. On the other hand, using $G_A(t) = G_R(-t)$, the second term vanishes, since 
\begin{eqnarray}
  & & \int dt_1 dt_2 \frac{1}{4}  G_R(t_1) G_A(t_2)  \text{sgn}(t_1 + t_2) \nonumber \\
  &=& \int dt_1 dt_2 \frac{1}{4}  G_R(t_1) G_R( - t_2)  \text{sgn}(t_1 + t_2) \nonumber \\
  &=& \int dt_1 dt_2 \frac{1}{4}  G_R(t_1) G_R(t_2)  \text{sgn}(t_1 - t_2) \nonumber \\
    &=& 0 ,
\end{eqnarray}
where we used the anti-symmetry of the integrand under the exchange of $t_{1,2}$. Hence, we have derived Eqn.~\ref{eq:identity_goal}. We comment on the validity of the substitution of the quantum FDT by its classical version at the end of Section~\ref{sec:high_T_lifetime}.

\subsection{$\tau_{\text{ph}}$ at low and high temperatures}

\label{sec:estimate_tau}

Here, we provide a simple derivation for the phonon lifetime in the limit of low and high temperatures in the disordered phase of the SB model. We assume that the system's parameters are chosen such that we are always far from the glass phase. To obtain an equation for the phonon lifetime, we consider the imaginary part of the retarded self-energy. Substituting $\hat{G}_K$ into $\hat{\Pi}_R$ in Eqn.~\ref{eq:Keldysh_sc_eqs} and taking the imaginary part of both sides reads 
\begin{equation}
    \text{Im}\left[\hat{\Pi}_{R}\left(\omega\right)\right] = -2v^2 \int \frac{d\omega'}{2\pi}  \mathcal{A}(\omega-\omega')\mathcal{A}(\omega')\coth\left(\frac{\beta\omega'}{2}\right).
    \label{eq:sc_for_tau_step_1}
\end{equation}
Here $\mathcal{A}(\omega) \equiv \text{Im}\hat{G}_R(\omega)$ is the spectral function. To proceed, we use the ansatz for the retarded propagator, given in Eqn.~\ref{retarded_ansatz}. It is then sufficient to consider the $\omega\to0$ limit of Eqn.~\ref{eq:sc_for_tau_step_1} in order to extract $\tau_{\text{ph}}$. The zeroth order in $\omega$ vanishes, and the leading term is given by 
\begin{equation}
    -\frac{2\omega}{\tau_{\text{ph}}} = -2v^2\omega \int \frac{d\omega'}{2\pi}  \partial_{\omega'}\mathcal{A}(\omega')\mathcal{A}(\omega')\coth\left(\frac{\beta\omega'}{2}\right),
    \label{eq:sc_for_tau_step_2}
\end{equation} 
We integrate the RHS by parts, use the fact that the spectral function decays at $|\omega|\to\infty$ and use the parity of the integrand to obtain
\begin{equation}
   \frac{1}{\tau_{\text{ph}}} = \frac{v^2\beta}{2} \int_0^{\infty} \frac{d\omega}{2\pi}  \left(\mathcal{A}(\omega)\text{csch}\left(\frac{\beta\omega}{2}\right)\right)^2.
    \label{eq:im_sc_for_tau}
\end{equation}

In the following, we will substitute the spectral function obtained from our ansatz (Eqn.~\ref{retarded_ansatz}) and solve for $\tau_{\text{ph}}$ at the high- and low-temperature limits. 



\subsubsection{High $T$}

\label{sec:high_T_lifetime}
High temperatures can be identified with the classical limit of the model in the following manner. In general, classical systems that obey Newtonian dynamics are invariant under the rescaling of the real-time coordinate $t\to\bar{t}=t/\sqrt{M}$, where $M$ is the mass of the classical degree of freedom. This implies that any time scale, and in particular $\tau_\text{ph}$, should be proportional to $\sqrt{M}$ at the classical limit. Furthermore, as this is a property of the equation of motion, we should be able to see this rescaling invariance at the level of the Keldysh saddle-point equations. It is easy to check that the Keldysh equations are invariant under this rescaling in frequency space (where $\omega \to \bar{\omega} = \sqrt{M}\omega$) if the temperature satisfies
\begin{equation}
    T\gg \omega_{\text{sf}},
    \label{eq:classical_T}
\end{equation}
where $\omega_\text{sf}$ is the characteristic frequency beyond which $\mathcal{A}(\omega)$ becomes negligible. In this limit, the rescaling invariance follows from the fact that the quantum FDT can be replaced with its classical version. In practice, Eqn.~\ref{eq:classical_T} holds when $T/\hbar$ is much larger than the largest frequency scale in the system. In terms of our ansatz, the classical limit of the model is identified with  $T\gg \sqrt{\Pi_0(T)}$. Recall that in Sec.~\ref{sec:real_time_Pi_0} and Sec.~\ref{sec:instability} we have derived an equation for $\Pi_0$ at high temperatures, whose solution is given by $\Pi_0(T) \approx \left(uT\right)^{1/2}$. Using this form, we may identify the classical limit with temperatures that satisfy $T\gg \hbar \Omega_u$. 
Note also that the criterion in Eqn.~\ref{eq:classical_T} is highly incompatible with the Planckian regime, where $\hbar \omega_\text{sf} \sim T$. 



We now proceed to estimate $\tau_{\text{ph}}$. In terms of our ansatz, the spectral function reads
\begin{equation}
    \mathcal{A}(\omega) = \frac{2\omega/\tau_\text{ph}}{\left(\omega^2 - \Pi_0\right)^2 + \left(2\omega/\tau_\text{ph}\right)^2 }.
    \label{eq:sf_high_T}
\end{equation}

Substituting Eqn.~\ref{eq:sf_high_T} in Eqn.~\ref{eq:im_sc_for_tau}, and using $T\gg \omega_\text{sf} \sim \sqrt{\Pi_0(T)}$ to expand the \text{csch} function to leading order, we have that
\begin{eqnarray}
  \tau_{\text{ph}} &=& 8 v^2T\int_0^{\infty}\frac{d\omega}{2\pi} \left(  \frac{1}{\left(\omega^2 - \Pi_0\right)^2 + \left(2\omega/\tau_\text{ph}\right)^2 }  \right)^2 \nonumber\\
                &\approx & \frac{1}{8} \frac{v^2T}{\Pi_0^2} \tau_\text{ph}^3
                \label{eq:tau_high_T_intermed_step}
\end{eqnarray}
where in the last line we have used the assumption that $\tau_\text{ph} \sqrt{\Pi_0} \gg 1$ to ignore subleading contributions in $1/\tau_\text{ph} \sqrt{\Pi_0}$. Finally, we substitute $\Pi_0(T) \approx \left(uT\right)^{1/2}$ into Eqn.~\ref{eq:tau_high_T_intermed_step} and obtain the high-$T$ phonon lifetime as given by Eqn.\ref{eq:tau_high_T}.


As for the consistency of this approximation and the one in section \ref{sec:real_time_Pi_0}: we have found that $\Pi_0 \approx \left(uT\right)^{1/2}$ for $T\gg \Omega_u$, which then implies that (see discussion below Eqn.~\ref{classical_lifetime_prop_to_sqrtM})
\begin{equation}
    \tau_{\text{ph}}^{-1} \ll \sqrt{\Pi_0}.
\end{equation}
Namely, the spectral function at high-$T$ is peaked around $\omega \approx \sqrt{\Pi_0}$ and its width is much smaller than $\sqrt{\Pi_0}$ In this case, one can indeed replace the quantum FDT by its classical version as the integrands in Eqn.\ref{eq:real_time_Pi_0_before_classical_FDT} and Eqn.\ref{eq:tau_high_T_intermed_step} are essentially supported in a frequency interval for which $\omega/T\ll 1$, as we have assumed in our self-consistency argument.

\subsubsection{Low $T$}

At low temperatures we expect the phonon lifetime to be exponentially long,  $\tau_\text{ph} \sim e^{\sqrt{\Pi_0(T\to 0)}/ \eta T}$, where $\sqrt{\Pi_0(T\to 0)}$ is the $T=0$ gap and $\eta$ is some numerical coefficient. 

Now, in the limit $T\ll \omega_\text{sf} \sim \Omega_0$, which corresponds to $\omega'\beta\gg1$ in the support of the integrand in Eqn.~\ref{eq:im_sc_for_tau}, we may approximate $\text{csch}(\omega'\beta/2) \approx 2e^{-\omega'\beta/2}$. Then, Eqn.~\ref{eq:im_sc_for_tau} is given by 
\begin{equation}
   \frac{1}{\tau_{\text{ph}}} = 2v^2\beta\int_0^{\infty}\frac{d\omega'}{2\pi}   \left(\mathcal{A}(\omega')e^{-\beta\omega'/2}\right)^2.
    \label{eq:low_T_step0}
\end{equation}
We proceed by using the ansatz Eqn.~\ref{eq:sf_high_T},  
\begin{equation}
   \frac{1}{\tau_{\text{ph}}} = 2v^2\beta\int_0^{\infty}\frac{d\omega'}{2\pi}   \left(\frac{2 e^{-\beta\omega'/2} \omega'/\tau_\text{ph}}{\left(\omega'^2 - \Pi_0\right)^2 + \left(2\omega'/\tau_\text{ph}\right)^2 } \right)^2.
    \label{eq:low_T_step1}
\end{equation} 
Note that if $\tau_\text{ph}\sqrt{\Pi_0},\tau_\text{ph}T\gg1$, the leading contribution to the integrand is coming from $\omega' \approx \sqrt{\Pi_0}$. We may therefore approximate the integral similarly to the high temperature limit and obtain  
\begin{equation}
       \frac{1}{\tau_{\text{ph}}} \approx \frac{2v^2\beta}{2\pi}\left(\frac{1}{\tau_\text{ph}} \right)\left(\frac{ e^{-\beta \sqrt{\Pi_0}} }{\left(2 \sqrt{\Pi_0}/\tau_\text{ph}\right)^2}\right).
\end{equation}
Finally, the phonon lifetime is given by Eqn.~\ref{eq:tau_low_T}.

\section{Numerical solution of the saddle-point equations}
\label{sec:numerical_methods}
Here we give some details on the numerical solution of the self-consistent equations in imaginary- and real-time for the single-branch model. The generalization of these methods to the MB model is straightforward with the use of the summation described in footnote~\ref{foot_note_implementation}.
\subsection{Imaginary-time}
\subsubsection{Disordered phase (replica diagonal)}
\label{imag_time_for_dis_phase}
We solve the saddle-point Eqs.~\ref{eq:G_Pi_diagonal} iteratively following the method of \cite{davison_thermoelectric_2017}, with small modifications to be specified ahead.
We first describe the unmodified algorithm. At the zeroth iteration step we use an initial condition $\hat{G}_0^{-1}(i\omega_n)$. The Matsubara frequencies are given by $\omega_n=2\pi\beta n$ where $-P_e \leq n \leq P_e$ such that the total number of sampling points is $P=2P_e + 1$. After the $j$th iteration step, we obtain $\hat{G}_{j}$, and substitute it into  
\begin{equation}
    \hat{\Pi}_{j+1}\left(i\omega_{n}\right) = \frac{v^{2}}{\beta}\sum_{\omega_n'}\hat{G}_j\left(i\omega_{n'}\right)\hat{G}_j\left(i\omega_{n}-i\omega_{n'}\right)-uG_j\left(0\right)
\end{equation} 
which we implement using MATLAB's \texttt{conv\char`_fft2} function package. The convolution function outputs a length $2P-1$ vector from which we take only the $P$ components that contain frequencies in the same window as $\hat{G}_j$. We then update the two-point function as 
\begin{equation}
    \hat{G}_{j+1}\left(i\omega_{n}\right) = (1-X) \hat{G}_{j}\left(i\omega_{n}\right) + X\frac{1}{\omega_n^2 + \Omega_0^2 - \hat{\Pi}_{j+1}\left(i\omega_{n}\right)},
    \label{eq:updating_scheme}
\end{equation}
where $X\in(0,1)$. 

After each iteration step we monitor the error $e_j\equiv \int |G_j - G_{j+1}|^2$ and if $e_j>e_{j-1}$ we update $X\to X/2$, as long as $X$ is larger then a minimal updating factor (to ensure convergence). The iteration procedure is terminated when $|e_j-e_{j-1}|<\epsilon$.  Usually we start with $X=1/2$, set the minimal updating factor to $X_{min} = 1/100$, and set $\epsilon = 10^{-14}$. Within the unmodified algorithm that we are currently discussing, we obtain the Green's function for a given $\Omega_0$ adiabatically by approaching it from above. That is, we start by solving for the Green's function with a large $\Omega_0$ (e.g. $\Omega_0/\Omega_v = 4$), with the initial condition of a free phonon. Then we slightly decrease $\Omega_0$ and use the previously obtained solution as the initial condition for the slightly decreased $\Omega_0$. In general, one may encounter multiple solutions for saddle-point equations as a function of $\Omega_0$. However, the solutions obtained with the above method are associated with the thermodynamically favorable saddle point of the disordered phase, as we have checked by comparing the free energies of the different solutions. 

We proceed to describe the modified algorithm. This modification enables us to approach thermodynamically unfavorable solutions of the disordered saddle-point equations, and conveniently solve the saddle-point equations in the glass phase. Instead of treating $\Omega_0$ as an input (together with $v,u$ and $\beta$), we fix $\hat{G}(i\omega=0)$ and then ask what $\Omega_0$ corresponds to such $\hat{G}(i\omega=0)$. To implement this we introduce $\delta\Pi \equiv \hat{G}(0)^{-1} $ and redefine the iteration step as follows,
\begin{equation}
    \hat{G}_{j+1}\left(i\omega_{n}\right) = (1-X) \hat{G}_{j}\left(i\omega_{n}\right) + X\frac{1}{\omega_n^2 + \delta\Pi - \tilde{\hat{\Pi}}_{j+1}\left(i\omega_{n}\right)},
    \label{eq:modified_updating_scheme}
\end{equation}
where $\tilde{\hat{\Pi}}_{j+1}\left(i\omega_{n}\right) \equiv \hat{\Pi}_{j+1}\left(i\omega_{n}\right) - \hat{\Pi}_{j+1}\left(0\right)$. Once the iteration process converges we can extract its corresponding $\Omega_0^2$, given by $\hat{\Pi}(0) + \delta\Pi$. Note that the correspondence is not single valued due to the existence of multiple saddle-point. The solutions with this approach are always obtained with a free initial condition. No adiabatic tuning is needed. 

The 1SRSB solution is obtained from the straightforward generalization of this method to the saddle-point Eqs.~\ref{eq:Pi_tilde_1SRSB}, and by setting $\delta\Pi = \frac{x}{m\beta g_{EA}}$. The conformally invariant solution is obtained by setting $\delta\Pi = \hat{\Pi}_\text{conf}(0)$ for a given $\beta$, where $\hat{\Pi}_\text{conf}(0)$ is given by the inverse of Eqn.~\ref{conformal_G} at $\omega_n=0$. We find that the conformally invariant solution of the disordered saddle-point equations, where $\Omega_0 = \Omega_\text{conf}$ and $\delta\Pi = \hat{\Pi}_\text{conf}(0)$, is not the thermodynamically favorable solution for this value of $\Omega_0$. Namely, we find that there exists $\delta\Pi > \hat{\Pi}_\text{conf}(0)$ that corresponds to $\Omega_{\text{conf}}$ and has lower free-energy. 

The computation of the free-energy density of the glass and disordered phases is done using Eqn.~\ref{eq:free_energy_glass} and Eqn.~\ref{eq:free_energy_dis}, respectively. The computation of the specific heat is done by computing the numerical derivative of the internal energy given by Eqn.~\ref{internal_E}.

\subsection{Real-time}

In this section we discuss the numerical solution of the real-time saddle-point equations and discuss the extraction of $\tau_{\text{ph}}$. The saddle-point equations are solved iteratively with an algorithm that is largely along the lines of \cite{song_strongly_2017}. 

The algorithm contains the following steps. We begin with an initial condition $\hat{G}_{R,0}(\omega_r)$ where $\omega_r$ are discretized real-frequencies, $\omega_r = \frac{2\pi}{t_{\text{max}}}r$  where $t_\text{max}$ denotes the maximal time such that $t_r = \frac{r}{P}t_{\text{max}}$, and $-P_e \leq r \leq P_e$ such that the total number of sampling points is $P=2P_e + 1$. After the $j$th iteration step we obtain $G_{R,j}(\omega_r)$. 
We then construct the Keldysh Green's function using Eqn.~\ref{eq:Keldysh_sc_eqs} \footnote{The zero-frequency component  $\hat{G}_{K,j}\left(\omega_{\frac{N+1}{2}}\right)$ is obtained by expanding the $\coth$ function and $\text{Im}\left[\hat{G}_{R,j}\right]$ to first order.},
\begin{equation}
    \hat{G}_{K,j}\left(\omega_r\right)=2i\coth\left(\frac{\beta\omega_r}{2}\right)\text{Im}\left[\hat{G}_{R,j}\left(\omega_r\right)\right].    
    \label{eq:numerics_G_K}
\end{equation}

To obtain the retarded self-energy we use MATLAB's \texttt{\char`fft} function to transform the Keldysh and retarded Green's functions to real-time, which we then pad by setting $G_{R,j}(t_{r'}) = 0$ for all $r'>\frac{P+1}{2}$ (in other words, we set $G_R(t)\equiv0$ for $t>t_{\text{max}}/2$). The retarded self-energy is then constructed in real-time according to Eqn.~\ref{eq:Keldysh_sc_eqs},
\begin{equation}
    {\Pi}_{R,j+1}\left(t_r\right) = iv^{2}{G}_{R,j}(t_r){G}_{K,j}(t_r) -\frac{iu}{2}G_{K,j}\left(t_{1}\right)\delta_{r,1}. 
    \label{eq:Pi_R_numerics}
\end{equation}

To finalize the iteration step we use MATLAB's \texttt{\char`ifft} function to obtain ${\hat{\Pi}}_{R,j+1}\left(\omega_r\right)$ and update the retarded Green's function as follows, 
\begin{equation}
    \hat{G}_{R,j+1}\left(\omega_{r}\right) = (1-X) \hat{G}_{R,j}\left(\omega_{r}\right) + X\frac{-1}{\omega_r^2 - \Omega_0^2 - \hat{\Pi}_{R,j+1}\left(\omega_{r}\right)},
    \label{eq:updating_scheme_real_time}
\end{equation}
where $X\in(0,1)$ is a constant updating factor. Convergence is reached when $|e_j - e_{j+1}|<\epsilon$ where $e_j = \int|G_{R,j}(t) - G_{R,j-1}(t)|^2 dt$. We choose $P_e = 2^{19}$, $X=0.1$, $\epsilon = 10^{-19}$ and the initial condition is always taken to be $-\hat{G}_{R,0}(\omega_r)^{-1} = (\omega_r+i\eta)^2 - \Omega_0^2$ where $\eta$ may be used as a tuning parameter (usually $\eta=1/3$). We tune $t_{\text{max}}$ to satisfy two self-consistency conditions $|G_R(t_{\text{max}}/2)|/|G_R(0)|\ll 1$ (or equivalently $t_{\text{max}} \gg \tau_{\text{ph}}$) and $|G_R\left(\omega_{P}\right)|/|G_R\left(\omega_{\frac{P+1}{2}}\right)|\ll 1$. We find that $t_{\text{max}} = C\beta$ with $C=\mathcal{O}(10^3)$ fulfill these conditions in a considerably large region in $(T,\Omega_0)$-space. Note, however, that this region is limited by low temperatures at which the phonon lifetime tends to increase exponentially, or high temperatures, where the $\tau_\text{ph}$ saturates to a constant as $\beta \to 0$. Fortunately, the temperature range at which the iteration process converges is sufficient for our needs. We find that the results depend only weakly on the choice of the tuning parameters $\eta,X,t_\text{max}$, giving uncertainty of a few percents in $\tau_\text{ph}$.

To verify this iteration procedure we use the spectral representation of the Matsubara Green's function,
\begin{equation}
    \hat{G}(i\omega_n) = \int_{-\infty}^{\infty} \frac{d\omega}{\pi} \frac{\mathcal{A}(\omega)}{\omega - i\omega_n},
    \label{eq:spectral_rep}
\end{equation}
to construct $G(\tau)$ out of $G_R(t)$. We find good agreement between the two, see Fig.~\hyperref[fig:linear_fit_demonstration]{\ref{fig:linear_fit_demonstration}a}.

\subsubsection{Phonon lifetime}

\begin{figure}[H]
    \centering
    \includegraphics[width=\columnwidth]{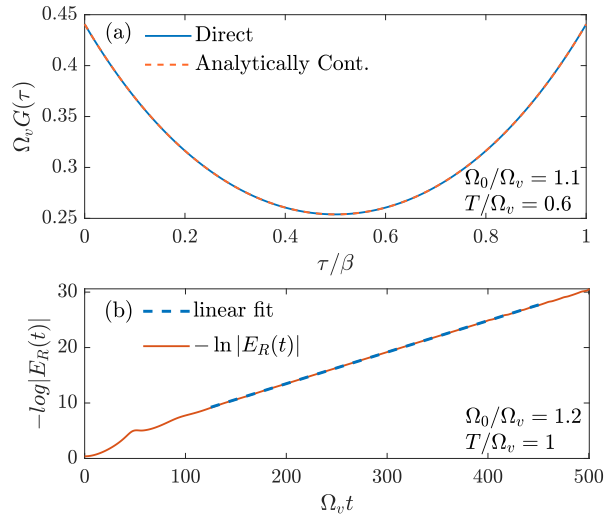}
  
    \caption{(a) Comparison between the imaginary time Green's function which we obtain by directly solving Eqs.~\ref{eq:G_Pi_diagonal} and the imaginary-time Green's function we obtain from the real-time data, using the spectral representation (see Eqn.~\ref{eq:spectral_rep}). (b) Example for a numerical extraction of the phonon lifetime of the SB model, where long- and short-times (right and left of the fitted interval, respectively) are disregarded. Here $u/\Omega_v^3 = 1.4$.}
        \label{fig:linear_fit_demonstration}
\end{figure}

The extraction of $\tau_{\text{ph}}$ is done as follows. Assuming that $G_R(t)\sim e^{-t/\tau_{\text{ph}}}$, we extract its envelope using MATLAB's \texttt{envelope} function (by extrapolating between the local maxima of $|G_R(t)|$). Denoting the output of this procedure $E_R(t)$, our assumption implies that $-\ln{(E_R(t))}$ at late-times is linear and its slope is equal to $1/\tau_{\text{ph}}$. In practice $\tau_{\text{ph}}$ is extracted by a linear fit in a time window $[t_1,t_2]$, where $t_1,t_2$ should be chosen such that $\tau_\text{ph} \ll t_1$ and $t_2\ll t_\text{max}$. Fig.~\hyperref[fig:linear_fit_demonstration]{\ref{fig:linear_fit_demonstration}b} demonstrates a typical fitting procedure where the long- and short-times parts are disregarded.

\end{document}